\documentclass[fleqn,usenatbib]{mnras}
\usepackage[T1]{fontenc}
\usepackage{ae,aecompl}

\usepackage{natbib}
\usepackage{comment}
\usepackage{color}

\usepackage{graphicx}	
\usepackage{amsmath}	
\usepackage{amssymb}	

\newcommand{\ok}{}
\newcommand{\bh}{}
\newcommand\scalemath[2]{\scalebox{#1}{\mbox{\ensuremath{\displaystyle #2}}}}

\title[UHECR from Virgo~A]{\ok{Search for ultra high energy cosmic rays from radiogalaxy Virgo~A}} 

\author[O. Kobzar et al.]{
Oleh Kobzar,$^{1}$\thanks{E-mail: Oleh.Kobzar@ifj.edu.pl}
Bohdan Hnatyk,$^{2}$
Volodymyr Marchenko,$^{3}$
Oleksandr Sushchov$^{1}$
\\
$^1$Institute of Nuclear Physics PAS, 152 Radzikowskiego Str., 31-342 Krak\'{o}w, Poland\\
$^2$Astronomical Observatory of Taras Shevchenko National University of Kyiv, 3 Observatorna Str., 04053 Kyiv, Ukraine\\
$^3$Astronomical Observatory, Jagiellonian University, 171 Orla str., 30-244 Krak\'{o}w, Poland
}

\date{Accepted 2019 January 1. Received 2018 December 18; in original form 2018 October 24}

\pubyear{2019}

\begin{document}
\label{firstpage}
\pagerange{\pageref{firstpage}--\pageref{lastpage}}
\maketitle

\begin{abstract}

Active galactic nuclei (AGNs) are considered to be one of the most appropriate sources of ultra high energy cosmic rays (UHECRs, $E \gtrsim 10^{18} \mathrm{~eV}$). Radiogalaxy Virgo~A (M87) in the centre of a cluster of galaxies Virgo Cluster (VC)  can be a prominent source of UHECRs. We investigate the possible contribution of Virgo~A and the VC to the flux of events with trans-GZK energies -- extremely high energy cosmic rays (EHECRs) -- from the recent Auger and Telescope Array  (TA)  data sets ($E > 52 \mathrm{~EeV}$ and $E > 57 \mathrm{~EeV}$, respectively). We simulate EHECR propagation from Virgo~A and the VC taking into account their deflections in galactic (GMF) and  extragalactic (EGMF) magnetic fields and show that there is no excess of EHECR arrival directions from Virgo A/VC images at different EHECR rigidities. By means of event-by-event analysis we recover the extragalactic arrival directions of EHECR events detected by Auger and TA for the representative set of nuclei H(p), He, N, Si, Fe, and find evidences of enhanced fluxes of N-Si-Fe EHECRs from the Local Filament and Hot/Cold Spot regions. The Local Filament with its enhanced magnetic field is an expected contributor to the UHECR flux as the closest to Earth last scattering centre, whereas Hot/Cold Spot region is a part of a larger arc-like spot, possibly created by the diffusively spreading jet of UHECRs, accelerated in the relativistic jet of Virgo A during a prominent nuclear outburst about 10 -- 12 Myr ago.
 
\end{abstract}




\section{Introduction}
\label{sec:introduction}

The origin of ultra high energy cosmic rays (UHECRs) with energies exceeding $\sim 10^{18} \mathrm{~eV}$ is still the subject of discussion \citep{2000RvMP...72..689N,2011ARA&A..49..119K,2014APh....53..120B,2014JCAP...10..020A}. While diffusive shock acceleration in Galactic sources (mainly in Supernova remnants) can be responsible for the observed flux of CR with $E \lesssim 10^{18} \mathrm{~eV}$ \citep{2013A&ARv..21...70B,2017PhRvL.119s1102N}, the list of potential sources of UHECRs includes a variety of extragalactic objects -- active galactic nuclei \citep[AGN;][]{1993A&A...272..161R,1995ApJ...454...60N,2009NJPh...11f5016D,2009ApJ...693..329F,2015APh....70...54F}, gamma-ray bursts \cite[GRBs;][]{1995PhRvL..75..386W,1995ApJ...453..883V,1999MNRAS.305L...6G,2003ApJ...592..378V,2012ApJ...753...69H,2015MNRAS.451..751G,2018PhRvD..97h3010Z}, starburst galaxies \citep[SBGs;][]{2014NuPhS.256..241M, 2016arXiv161000944B}, newborn millisecond pulsars and magnetars \citep{2003ApJ...589..871A,2012ApJ...750..118F,2015JCAP...08..026K,2016ApJ...826...97P}, tidal disruption events (TDEs) in supermassive black hole gravitation fields \citep{2009ApJ...693..329F,2014arXiv1411.0704F,2017PhRvD..96f3007Z} etc.  

The problem of identifying the sources of UHECR is seriously complicated by roughly isotropic distribution of UHECR arrival directions on the celestial sphere and by a lack of their reliable correlations with the positions of potential sources. For example, Auger found only global (dipole) anisotropy in the arrival directions of UHECRs with $E > 8 \mathrm{~EeV}$ with an amplitude of $6.5^{+1.3}_{-0.9} \%$ and directions $RA = (100 \pm 10)^{\circ}$, $Dec = -(24^{+12}_{-13})^{\circ}$ ($5.2\sigma$ CL) \citep{2017Sci...357.1266P,2018arXiv180508220A}. 
The TA surface detector data for the sky regions inside/outside $\pm 30^{\circ}$ from the supergalactic plane ("on-source/off-source" regions) reveals different spectra at trans-GZK energies from these regions:  the off-source spectrum has an earlier break \citep{2017arXiv170704967A}. Joint analysis of Auger and TA data shows additional signatures of global anisotropy: a harder (with a later break) northern hemisphere ($Dec > 24.8^{\circ}$) spectrum \citep{2018arXiv180107820A} and gradual decrease of the ratio of Auger and TA UHECR ($E > 10 \mathrm{~EeV}$) fluxes with the increase of energy in the common declination band $-15.7^{\circ} \leq Dec \leq +24.8^{\circ}$ \citep{2018arXiv180101018T}.

At intermediate-scale 5 yr TA data of arrival directions of $E > 57 \mathrm{~EeV}$ UHECRs revealed a "hot spot" \ok{with a statistical significance of $5.1 \sigma$}: the cluster of events within $20^{\circ}$ radius circle around $RA = 146.7^{\circ}$, $Dec = 43.2^{\circ}$ \citep{2014ApJ...790L..21A}. More recent analysis of 9 yr TA data  \citep{insert1} confirmed a hot spot with $3\sigma$ global significance for $25^{\circ}$ oversampling, together with the presence of a "cold spot": the deficit of $10^{19.2} \leq E \leq 10^{19.75} \mathrm{~EeV}$ events \citep{insert2}.

Meantime, no signs of UHECR excess from potential close sources, including the Virgo Cluster (VC), was found in the TA and Auger data \citep{insert1}.

Determination of chemical (mass) composition of UHECRs is of principal value for the search for their sources, but experimental data is still controversial. According to the recent Auger data \citep{2017arXiv170806592T,2017arXiv171007249T}, the UHECR flux is predominantly composed of light (H, He) nuclei at around $2 \mathrm{~EeV}$, but the average atomic mass increases up to $40 - 50 \mathrm{~EeV}$ to He -- N values. In the energy bins for $E > 50 \mathrm{~EeV}$ the observed stop  of the primary mass increase may be affected by statistical and systematic uncertainties.

According to the recent TA data \citep{2018ApJ...858...76A}, pure proton composition with the QGSJetII-04 hadronic interaction model is still compatible with the TA hybrid data in $18.2 \leq \lg E \leq 19.9$ investigated range at the $95 \%$ confidence level, taking into account some systematic shifting of shower maximum depth $X_{\mathrm{max}}$. 
The average atomic mass of UHECRs in the TA data shows no significant energy dependence in the whole $18.2 \leq \lg E \leq 20.0$ range and  is estimated as $\langle \ln A \rangle = 1.52 \pm 0.08 \mathrm{~(stat)} \pm 0.36 \mathrm{~(syst)}$ \citep{2018arXiv180803680T}.

Both observable arrival directions and chemical composition of UHECRs do not correspond to real sources positions and injected composition due to deflection in the GMF and the EGMF and to photodisintegration of nuclei correspondingly. These factors are minimal for extremely high energy cosmic rays (EHECRs, $E \geq 5 \times 10^{19} \mathrm{~eV}$) from nearby (close to Earth) sources. Therefore, event-by-event analysis of EHECRs from nearby potential sources, particularly, AGNs, is a promising way of finding out specific physical sources of UHECRs. Such a search was carried out for Centaurus~A, the nearest AGN to the Milky Way, located at a distance of $3.8 \mathrm{~Mpc}$ and surrounded by the localized group of UHECR events \citep{2008APh....29..188P,2010PrPNP..64..363F}. With respect to the possible deflections of charged CR particles in the GMF it was shown that at least some of these events could really originate from Centaurus~A \bh{\citep{2012KPCB...28..270S,2013JCAP...01..023F,2014A&A...562A..12P,2014ApJ...783...44F,2017arXiv170608229W,2017arXiv170905766F,2018MNRAS.479L..76M,2018MNRAS.481.4461F,2019MNRAS.482.4303M}}. 
\bh{Once more, \cite{2018JCAP...02..036E} propose Centaurus~A as the main source of heavy nuclei dominating in the EHECR flux.}

\bh{The expected contribution of UHECRs from the other brightest radio galaxies of the Local Universe to the observed flux was investigated for the cases of Fornax~A by \cite{2018MNRAS.479L..76M,2018JCAP...02..036E,2019MNRAS.482.4303M}, and Centaurus~B by \cite{2009ApJ...693.1261M,2018arXiv181101108F}, IC310 by \cite{,2017APh....89...14F}.}
  
In this paper we investigate a possible contribution of another close AGN, Virgo~A, into the observed EHECR flux.  Virgo~A is the second nearest AGN located at a distance of $16.5 \mathrm{~Mpc}$ \citep{2013AJ....146...86T}. 
According to contemporary estimations, it could be a prominent UHECR source \citep{2009JCAP...01..033D,2016ApJ...830...81F,2018JCAP...02..036E}.

In several recent works efforts aimed at the search for UHECR originated in Virgo~A were undertaken \citep{2011PhRvD..83h3002G,2015arXiv150909033S,2015ICRC...34..503K}.
Unlike Centaurus~A case, a zone of UHECR avoidance is observed in this sky region.
If Virgo~A is really an UHECR source, observed deficit of events can be explained by significant influence of the GMF and the EGMF, transient nature of the source and photodisintegration (for He nuclei).
On the other hand, the distance to Virgo~A  is too short for the GZK cut-off \citep{1966PhRvL..16..748G,1968CaJPS..46..617K} to be efficient even for the proton component.
Therefore, in order to search for EHECRs from Virgo~A in the observed flux (231 Auger events with $E > 52 \mathrm{~EeV}$ \citep{2015ApJ...804...15A} and 72 TA events with $ E > 57 \mathrm{~EeV}$ \citep{2014ApJ...790L..21A}), we take into account the GMF as well as the EGMF. For both regular and stochastic GMF we use JF12 model \citep{2012ApJ...757...14J,2012ApJ...761L..11J}.
The EGMF distribution follows the large scale structure (LSS) of matter density distribution, and in our Local Universe (LU), especially in the Virgo Supercluster (VSC), the main structures that determine the conditions of EHECR propagation are the VC, the Local Filament (LF) and voids around them  \citep{2003ApJ...596...19K,2007A&A...476..697E,2014Natur.513...71T}. Virgo~A is a promising candidate for a transient UHECR source with effective UHECR acceleration during the giant AGN flares \citep{2009ApJ...693..329F} alike one 12 Myr ago  \citep{2017ApJ...844..122F}, and we consider also EHECR signature of such a Virgo~A flare. 

This paper is organized as follows.
In Section \ref{sec:lss} we give a short review of the Large scale structure and intergalactic magnetic field in the VSC.
In Section \ref{sec:CRfromVC} we describe the numerical method we used and investigate the propagation of EHECR from the VC in the void EGMF and in the GMF.
In Section \ref{sec:filamentmodel} the influence of the LF on the EHECRs from Virgo~A and the VC is considered.
The signatures of the jet-accelerated EHECR in the giant flare event in Virgo~A are analysed in Section \ref{sec:jetmodel}.
Discussion and conclusions are presented in Section \ref{sec:DisCon}.

\section{Large scale structure and magnetic fields in the Virgo Supercluster}
\label{sec:lss}

LSS -- non-uniform cosmological matter density distribution -- affects UHECR propagation in two ways: by connected with baryonic matter intergalactic magnetic fields and by non-uniform UHECR sources distribution. In the LU within $\sim 100 \mathrm{~Mpc}$, where Laniakea Supercluster dominates \citep{2014Natur.513...71T}, and even within the VSC all components of the LSS, which contribute to generation and propagation of UHECRs, are present: voids, sheets, filaments, clusters of galaxies and galaxy superclusters \citep{2007A&A...476..697E}. The UHECR flux from Virgo~A and the VC should be considerably influenced by these structures.

\subsection{Large scale structure and the Virgo Supercluster}

The VSC is a typical poor "spider"-type  supercluster with one rich cluster of galaxies, the VC (with Virgo~A (M87) as its central galaxy)  in its centre, surrounded by $\sim 25 \mathrm{~Mpc}$ halo of filaments, where poorer clusters and groups of galaxies reside \citep{2007A&A...476..697E}.
 
In one of such filaments, called the LF, which extends from the VC to the Fornax Cluster, the Local Group of galaxies, that includes our Galaxy Milky Way, Andromeda galaxy (M31) and a set of dwarf galaxies, is located \citep{2003ApJ...596...19K}, \citep{2013AJ....146...86T,2015MNRAS.452.1052L,2016MNRAS.458..900C}. Both the VC and the LF lie in the Supergalactic plane, determined by preferential matter distribution in the LU.

Supergalactic coordinates of Virgo~A at a distance $D = 16.5 \mathrm{~Mpc}$ are the following: $SGL = 102.88^{\circ}$, $SGB = -2.35^{\circ}$, $SGX,~SGY,~SGZ = (-3.67;~+16.07;~-0.67) \mathrm{~Mpc}$. As  discussed later, the relativistic $~30''(\sim 2.3 \mathrm{~kpc})$ projected jet of Virgo~A is a promising source of UHECRs as well. The position angle of the jet axis $PA = 290^{\circ}$ \citep{2013ApJ...774L..21M}. According to \citep{2018ApJ...855..128W} we assume that the viewing angle for the jet is $17^{\circ}$ and the intrinsic (apparent)  jet opening angle  is $\theta \simeq 2(7)^{\circ}$. The point at the closest distance from the jet axis to our Galaxy $D_{\mathrm{min}} = 4.8 \mathrm{~Mpc}$ has coordinates $SGX,~SGY,~SGZ = (+3.13;~+0.59;~-3.60) \mathrm{~Mpc}$.

The LF has nearly cylindrical form with radius $R = 2 \mathrm{~Mpc}$ and  direction $SGL = 124^{\circ}$, $SGB= -2.4^{\circ}$ (about $22^{\circ}$ from Virgo~A direction) \citep{2015MNRAS.452.1052L,2016MNRAS.458..900C}.

In the Local Group region the LF shows up a disk-like matter density enhancement with radius of $\sim 4 \mathrm{~Mpc}$ and thickness of $\sim 2 \mathrm{~Mpc}$ -- the Local Sheet, that comprises, besides the Local Group, the NGC5128 (Centaurus~A) Group and the M81 Group (with a potential UHECR source starburst galaxy M82). The Local Sheet is practically the south wall of the Local Void -- a giant $\sim 45 \mathrm{~Mpc}$ void north to the Local Sheet \citep{2003A&A...398..479K,2013AJ....146...86T,2015MNRAS.452.1052L}.

In the LU the VSC is a part of the Laniakea Supercluster, together with  the Hydra-Centaurus Supercluster and the Great Attractor region dominated by the Norma Cluster (ACO 3627 or Abell 3627), the Pavo-Indus Supercluster and the Southern Supercluster \citep{2014Natur.513...71T,2017ApJ...845...55P}. The Laniakea Supercluster and the Perseus-Pisces Supercluster are two dominant concentrations of galaxies in the LU \citep{2014Natur.513...71T}, connected with a few V(velocity)-web filaments, and the VSC is a part of one of them -- Centaurus-Virgo-PP Filament or "Virgo strand" \cite{2017ApJ...845...55P}. 

\begin{figure}
\centering
\includegraphics[width=1.0 \linewidth, clip]{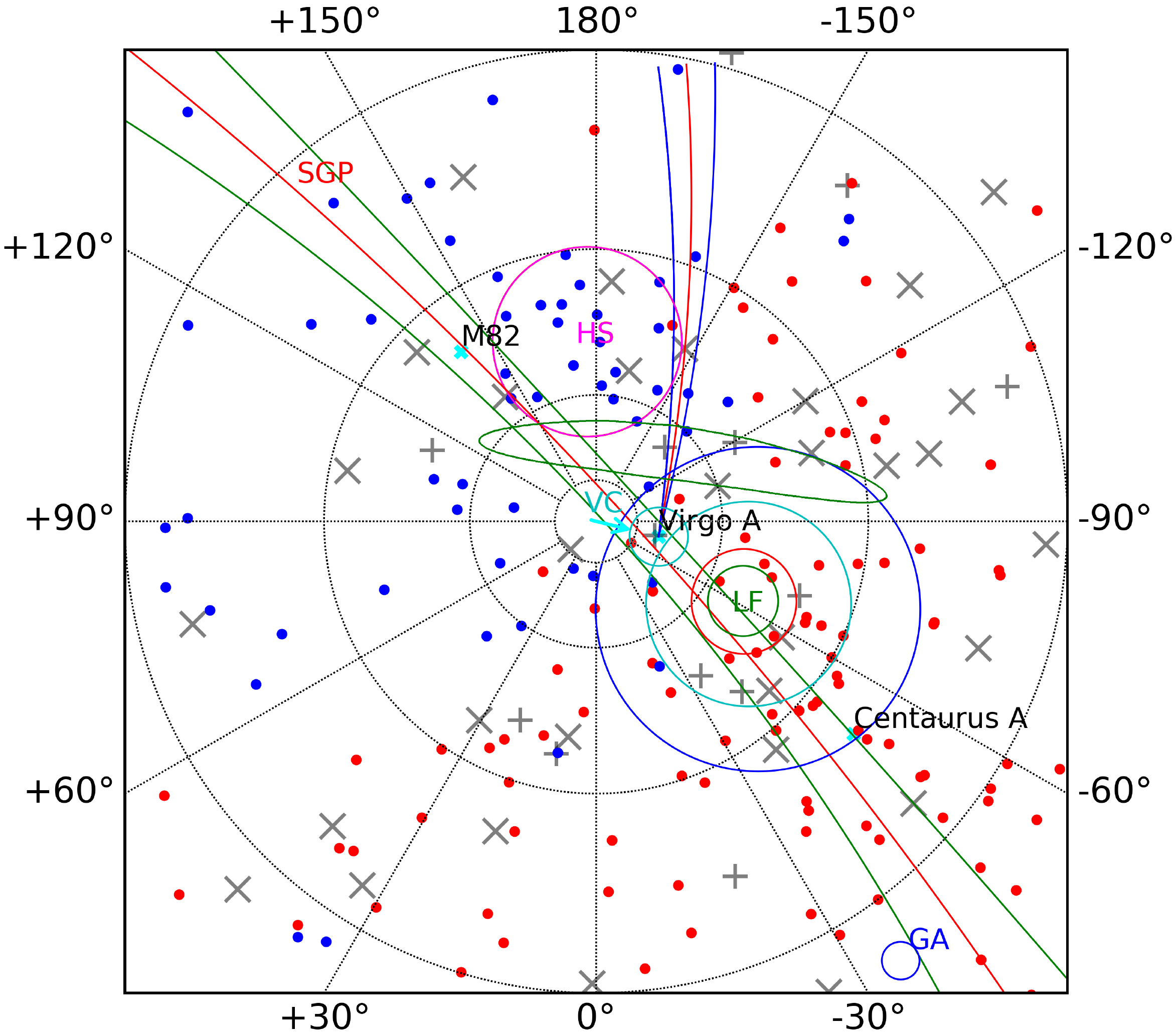}
\caption{The Virgo Supercluster structure. Galaxies Virgo A with a prolonged jet cone (red with blue boundaries) and jet-induced spot (green ellipse), Centaurus A and M82, together with the Supergalactic plane (SGP, red with green boundaries), the Virgo Cluster (VC, azure circle), the Great Attractor (GA) and the Local Filament (LF) are presented. Concentric circles around the LF represent the LF boundaries at distances 3 Mpc (blue), 5 Mpc (azure), 10 Mpc (red) and 15 Mpc (green) from the Earth. Red and blue points are Auger and TA events correspondingly. The Hot Spot (HS) in the TA data is shown with a pink circle. Crosses and pluses are the IceCube neutrino shower and track events correspondingly. The galactic coordinates with the North Galactic Pole in the center are used.}
\label{fig:vs}
\end{figure}

The LSS components are presented in Fig. \ref{fig:vs}

\subsection{Intergalactic magnetic fields in the Virgo Supercluster}
\label{sec:IGMF_VS}

Observed high level of isotropy for the UHECR intensity indicates a significant impact of intergalactic magnetic fields (IGMF) on their propagation, but the origin, spatial distribution and evolution of such fields are still unresolved \citep{1994RPPh...57..325K,2008ApJ...676...70K,2008Sci...320..909R,2013A&ARv..21...62D}. 
Till now only $\sim 1 \mathrm{~\mu G}$ magnetic fields in the central parts of clusters of galaxies are experimentally measured \citep{2012A&ARv..20...54F}, and only the upper limit (density-weighted) magnetic field of 0.03 (0.13) $\mathrm{~\mu G}$ in filaments at $z = 0$ together with the limit on the primordial magnetic field $< 1.0 \mathrm{~nG}$ are inferred in \citep{2017MNRAS.468.4246B} from the cross-correlation of the $2.3 \mathrm{~GHz}$ S-PASS survey with constrained magnetohydrodynamics (MHD) local cosmic web simulations. 
Similar $\sim$ nG upper limit on the smoothed over 1 Mpc primordial (void) magnetic fields is obtained in \citep{2016PhRvL.116s1302P} with the use of RM data from extragalactic sources and from Planck data combined with the South Pole Telescope CMB polarization measurements ($B_{1\mathrm{Mpc}} < 1.2 \mathrm{~nG}$ at 95\% C.L) \citep{2017PhRvD..95f3506Z}. 
Meantime, suppression of the low-energy (GeV) secondary gamma ray emission from TeV-emitting blazars suggests the lower limit on the IGMF in voids $B_{\mathrm{void}} \geq 3\times 10^{-16} \mathrm{~G}$ for the correlation length $\lambda_B \geq 0.1 \mathrm{~Mpc}$ \citep{2010Sci...328...73N}.

The main characteristics of the IGMF in the LU can be derived from the constrained simulations of its evolution that reproduce the observed LSS elements -- the VSC and the VC, the LF, the Local Group etc. \citep{2003ApJ...596...19K,2005JCAP...01..009D,2014MNRAS.445.3706V,2016MNRAS.458..900C,2018MNRAS.475.2519H}. Simulations of the common MHD evolution of matter and primordial or/and astrophysical dynamo-created cosmological magnetic fields, constrained by the LU observational data, including the observational data on values/upper limits of IGMFs in clusters/voids, result in the magneto-genesis scenario-dependence of the expected parameters of the IGMF in the LU and, especially, in the VSC. In the case of primordial magnetic field amplification  model  with initial volume-filling uniform $B_{0} \sim 10^{-9} \mathrm{~G}$ magnetic field, the present time field in a void is of the same value $B_{\mathrm{void}} \sim B_0 \sim 10^{-9} \mathrm{~G}$, of the order of $B_{\mathrm{fil}} \sim (2-5)\times 10^{-9} \mathrm{~G}$ in sheets/filaments and of the order of $B_{\mathrm{cl}} \sim 10^{-7} \mathrm{~G}$ in clusters of galaxies. Meantime, in the case of astrophysical models, where magnetic fields are seeded by stellar activity and galactic winds as well as by the outflows from the active galactic nuclei, the IGMF distribution is more contrasting with $B_{\mathrm{void}} \sim 10^{-15} - 10^{-14} \mathrm{~G}$ in voids, $B_{\mathrm{fil}} \sim 10^{-13} - 10^{-11} \mathrm{~G}$ in sheets/filaments and $B_{\mathrm{cl}} \sim 3 \times 10^{-8} - 3 \times 10^{-7} \mathrm{~G}$ in clusters of galaxies \citep{2014MNRAS.445.3706V,2016MNRAS.462.3660H,2018MNRAS.475.2519H}.

\subsection{Modelling of the Galactic magnetic field}
\label{subsec:GMF}

\bh{Galactic magnetic field with a characteristic amplitude of a few $\mu \mathrm{G}$ and of coherence length of a few kpc (regular fields of spiral arms etc.)/several tens of pc (random component of disc and halo) can considerably deflect the trajectories of even EHECRs, especially in the case of heavy nuclei. 
The expected deflection is strongly anisotropic, since extragalactic EHECRs with arrival directions from the circumpolar Galactic regions  (Virgo~A, Fornax~A etc.) interact mainly with the halo component, covering the distance of the order of Galaxy's radius, whereas EHECRs with arrival directions from the Galactic centre region  (Centaurus~A, Centaurus~B etc.) interact with the several times stronger regular component passing mostly along the Galactic disc diameter \citep{2008PhyS...78d5901F}.
}

Several GMF models were proposed during the last decades \citep{2007APh....26..378K,2011ApJ...738..192P,2012ApJ...757...14J,2012ApJ...761L..11J}. We use here the GMF model of \citep{2012ApJ...757...14J,2012ApJ...761L..11J} (JF model), based on the largest observational dataset.

According to JF model, regular GMF consists of spiral disk field and halo field, which includes toroidal and poloidal components. Unlike earlier models, in which poloidal field was treated as a galactic dipole, in JF model this component is represented as having an X-like profile in the vertical plane. 
Random GMF, in turn, is described as consisting of two parts: "striated" field, oriented along the large-scale regular field  but varying in strength and sign on a small scale, and actually random field, varying also in orientation. Both of these components are overlapped upon the large-scale regular field.
The average GMF magnitude is of the order of several  $\mathrm{\mu G}$ for the regular as well for the random components. We use analytical approximations of magnetic field characteristics, presented in \citep{2012ApJ...757...14J,2012ApJ...761L..11J}, in numerical calculations for both regular and random components.

\subsection{Influence of random magnetic fields on EHECR propagation: analytical estimations}
\label{subsec:RanMF}

In our search for EHECRs from the VC we limit ourselves with the case of their pre-diffusive (ballistic) propagation, with the maximum deflection angle \ok{$\theta \lesssim 90^\circ$}. The influence of random fields in this case can be described in terms of diffusion process in the particle momentum space. It looks like a small-angle random  scattering around the main trajectory in the regular field, that allows to calculate the average angular EHECR deflection $\Theta_{\mathrm{rms}}$ and the corresponding time delay $\tau_{\mathrm{rms}}$ \ok{(in years)} in the random field $B_{\mathrm{rms}}$ with the coherence length $l_{\mathrm{c}}$ \citep{1996ApJ...472L..89W,2003PhRvD..68d3002S} analytically: 

\begin{eqnarray}
\begin{aligned}
	& \Theta_{\mathrm{rms}}(E,L) = \sqrt{\frac{2}{9}} \left(\frac{Ze}{E}\right)
    B_{\mathrm{rms}} \sqrt{l_{\mathrm{c}} L} \\
    & \scalemath{0.95}{\simeq 0.8^\circ Z \left( \frac{E}{10^{20} \mathrm{eV}} \right)^{-1}
    \left( \frac{L}{10 \mathrm{Mpc}} \right)^{\frac{1}{2}}
    \left( \frac{l_{\mathrm{c}}}{1 \mathrm{Mpc}} \right)^{\frac{1}{2}}
    \left( \frac{B_{\mathrm{rms}}}{10^{-9} \mathrm{G}} \right) }
   	\label{eq:theta-ran}
\end{aligned}
\end{eqnarray}
and
\begin{eqnarray}
\begin{aligned}
	& \tau_{\mathrm{rms}}(E,L) = L \Theta^2_{\mathrm{rms}}(E,L)/4c \\
    & \scalemath{0.93}{\simeq 1500 Z^2 \left( \frac{E}{10^{20}eV} \right)^{-2}
    \left( \frac{L}{10 \mathrm{Mpc}} \right)^2
    \left( \frac{l_{\mathrm{c}}}{1 \mathrm{Mpc}} \right)
    \left( \frac{B_{\mathrm{rms}}}{10^{-9}G} \right)^2 }
	\label{eq:tau-ran}
\end{aligned}
\end{eqnarray}
where $L$ is the full distance covered by a CR particle on its way from the source to the observer. 

Within JF model the analytical approximation of the random GMF includes different random field strength $B_{\mathrm{rng,Gal}}$ in different arms \citep{2012ApJ...761L..11J}, and under the assumption of the common field coherence length $l_{\mathrm{c,Gal}}$ its impact can be calculated using the modified formula:
\begin{equation}
	\Theta_{\mathrm{rms,Gal}}^{2} = \frac{2}{9} \left( \frac{Ze}{E} \right)^{2}  
	l_{\mathrm{c,Gal}} \int B_{\mathrm{rms,Gal}}^{2} d(ct)~.
	\label{eq:scat-theta-Gal}
\end{equation}

Resulting scattering and time delay caused by random GMF and EGMF components are then
   \begin{equation}
	\Theta = \sqrt{\Theta_{\mathrm{rms,Gal}}^{2} + \Theta_{\mathrm{rms,EG}}^{2}}
	\label{eq:scat-theta}
\end{equation}
and
\begin{equation}
	\tau = (L_{\mathrm{Gal}}\Theta_{\mathrm{rms,Gal}}^{2} + L_{\mathrm{EG}}\Theta_{\mathrm{rms,EG}}^{2})/4c~.
	\label{eq:scat-tau}
\end{equation}

\subsection{EHECR acceleration in Virgo A and the VC}
\label{subsec:acceleration}

\cite{1984ARA&A..22..425H} showed that the maximum energy of CR in an accelerator of the size $L$ and magnetic field $B$ is restricted by a confinement condition on Larmor radius $r_{\mathrm{L}}$ of a particle with energy $E$ and electric charge $Ze$: $r_{\mathrm{L}} = E/ZeB < L$ or $E_{\mathrm{max}} = ZeBL$  (or for maximum rigidity (${\cal R} = E/Ze$) ${\cal R}_{\mathrm{max}} = BL$).
For acceleration in magnetized plasma, moving with velocity $v = \beta c$ , where $c$ is the speed of light, the corresponding electric field is of the order of ${\cal E} \sim \beta B$, and Hillas condition may be strengthened to $E_{\mathrm{max}} = Ze{\cal E}L = Ze \beta BL$.
In the Virgo~A case UHECR acceleration is expected in the radiolobes (up to 80 kpc in diameter) at the shock fronts created by interaction of a powerful enough (kinetic power $P_{\mathrm{kin}} \sim 10^{44} \mathrm{~erg~s^{-1}}$) jet \citep{2000ApJ...543..611O, 2006MNRAS.370..981S} with circumnuclear and interstellar  medium.
Both relativistic ($\beta_{\mathrm{jet}} \geq 0.9$, $B\sim 0.01 \mathrm{~G}$) pc-scale and subrelativistic ($\beta_{\mathrm{jet}} \leq 0.5$, $B \sim 300~\mu\mathrm{G}$) kpc-scale jets \citep{2006MNRAS.370..981S} are suitable regions for UHECR acceleration up to $E_{\mathrm{max}} \sim Z \times 10^{19} \mathrm{~eV}$.
As shown by \cite{2018MNRAS.473.2364B}, the first-order Fermi mechanism of diffusive shock acceleration provides the highest particle energy in the cases of nonrelativistic or mildly relativistic shocks. 
In Virgo~A as a FRI radiogalaxy these conditions are fulfilled in the radiolobes around kiloparsec-scale jets.
It has been shown \citep{1995PhRvL..75..386W,2000PhST...85..191B}, that from magnetic power of $\sim 1$  steradian-collimated jet of size $L$: 
$P_{B} \sim  L^{2}\beta_{\mathrm{jet}} c (B^{2}/8\pi)=\epsilon_{B} P_{\mathrm{kin}}$, 
where $\beta_{\mathrm{jet}}c$ is the jet velocity, and a fraction of magnetic power in kinetic one $\epsilon_{B} < 1$, it follows that (taking into account $ {\cal R} = \beta_{\mathrm{jet}}BL$) $ P_{\mathrm{kin}} \geq P_{B} = (\beta_{\mathrm{jet}}c/8\pi)(E/Ze)^{2} = 1.3 \times 10^{44}(\beta_{\mathrm{jet}})^{-1} ({\cal R}/10^{19} \mathrm{~V})^{2} \mathrm{~erg~s^{-1}}$, 
i.e., Virgo~A jet with $P_{\mathrm{kin}} \sim 10^{44} \mathrm{~erg~s^{-1}}$ and bulk jet velocity $\beta_{\mathrm{jet}} = 0.1$  can provide EHECRs with ${\cal R}_{\mathrm{max}} \sim 10^{19} \mathrm{~V}$ or $E_{\mathrm{max}} \geq 5 \times 10^{19} \mathrm{~eV}$ for nuclei with $Z > 6$ 
\citep{2018MNRAS.479L..76M,2019MNRAS.482.4303M}.

Additional constraints on maximum energy of UHECRs can follow from limiting of acceleration time 
$\tau_{\mathrm{acc}} \approx 3 r_{\mathrm{L}}/c \beta_{\mathrm{sh}}^{2} \approx 9\times 10^{5} {\mathrm{~yr}}$ \citep{2018JCAP...02..036E} by dynamical time 
$\tau_{\mathrm{dyn}} \geq 2\times 10^{6} {\mathrm{~yr}}$ 
(with the minimum value  during the outburst period \citep{2017ApJ...844..122F}), 
and escaping time downstream of the shock of radius $L_{sh}$: 
$\tau_{\mathrm{esc}} \approx L_{\mathrm{sh}}/c \beta_{\mathrm{sh}} \approx 3 \times 10^{5} {\mathrm{~yr}}$ \citep{2018JCAP...02..036E}.

Here we use one velocity scale $\beta_{\mathrm{jet}} = \beta_{\mathrm{sh}} = 0.1$, magnetic field scale $B = 30~\mu\mathrm{G}$, shock radius scale $L_{\mathrm{sh}} = 10 \mathrm{~kpc}$ and radiolobe radius $R_{\mathrm{lobe}} = 40 \mathrm{~kpc}$ as an accelerator size scale in case of acceleration on multiple shocks \citep{2019MNRAS.482.4303M}. As one can see, time scales of acceleration and escaping are similar. Other energy loss time scales (adiabatic, synchrotron, photomeson, photodisintegration) are considerably larger.

\subsection{Survival of EHECR nuclei on the way from the Virgo Cluster to the Earth}
\label{subsec:mfp}

Detected by the TA ($E > 57 \mathrm{~EeV}$) and  Auger ($E > 52 \mathrm{~EeV}$) EHECR events, considered in this paper, on their way to Earth can not be only deflected by the IGMF and the GMF, but also suffer from energy losses, and some of them can change their chemical composition via the photo-disintegration due to interaction with CMB and (mainly IR) Extragalactic  Background Light (EBL).  

The interaction path length for $E > 50 \mathrm{~EeV}$ protons is determined by interaction with CMB via  pair-production (starting at $E_{\mathrm{pp}} \sim 10^{18} \mathrm{~eV}$ per nucleon or at nucleus of atomic mass $A$ energy $E_{\mathrm{N}} = AE_{\mathrm{pp}}$) and photo-pion production (starting at GZK-cut-off energy $E_{\mathrm{GZK}} = 5 \times 10^{19} \mathrm{~eV}$ per nucleon), which results in the GZK horizon (energy loss length) $\sim 1000/100 \mathrm{~Mpc}$ for $E_{\mathrm{p}} = 5 \times 10^{19}/10^{20} \mathrm{~eV}$. The flux of $E > 50 \mathrm{~EeV}$ nuclei, apart from energy losses due to pair-production and chemical composition changes due to photodisintegration on far infra-red photons (low energy EBL) both at $E_{\mathrm{N}} > AE_{\mathrm{pp}}$, suffers mainly from photo-disintegration on CMB photons (at energies $E_{\mathrm{N}} > 4AE_{\mathrm{pp}}$) \citep[see, i.e.][and reference therein]{2017arXiv170708471A}. It means that energy losses of $E > 50 \mathrm{~EeV}$ protons on the LU scale ($\le 100 \mathrm{~Mpc}$) can be neglected. Meantime, even for He nuclei the total mean free path (energy losses and photodisintegration) is only $\sim 1/1 \mathrm{~Mpc}$ for $\lg E(\mathrm{eV}) = 19.7/20$ -- we do not expect to detect He nuclei accelerated in Virgo~A. C-N-O  group nuclei from Virgo~A with mean free path $\sim 46, 20, 15.3/3.3, 1.5, 1 \mathrm{~Mpc}$ for $\lg E(\mathrm{eV}) = 19.7/20$ \citep{2017arXiv170608229W} will suffer from photodisintegration only at the highest energies. 
\ok{Protons and heavy nuclei such as Si and Fe have the total mean free path larger than Virgo~A distance ($16.5 \mathrm{~Mpc}$) even for $\lg E(\mathrm{eV}) = 20$} \citep{2018JHEAp..17...38A} 

An unavoidable by-product of UHECR energy losses  is cosmogenic neutrino production \citep{1969PhLB...28..423B} via photo-pion production and subsequent charged pion and muon decay ($10^{15}-10^{18}$ eV neutrino), and via beta-decay of neutrons and nuclei produced by photo-disintegration ($10^{14}-10^{16}$ eV neutrino) \citep[][and reference therein]{2017arXiv170708471A}. Neutrinos are not deflected by magnetic fields and thus provide additional valuable information about the sources and trajectories of EHECRs. For analysis of neutrino events from the Virgo~A region (Fig. \ref{fig:vs}) we use the Ice Cube Collaboration observational data \citep{insert3,2018arXiv180101854T}.

\section{Propagation of UHECR from the VC in the Local Universe}
\label{sec:CRfromVC}

\subsection{Numerical calculation: Method}
\label{subsec:method}

According to the discussion in subsection \ref{subsec:mfp} we modelled the UHECR propagation neglecting energy losses. In the case of ultra relativistic rest mass $m_{o}$ particles with energy $E \gg m_{0}c^{2}$ and Lorentz-factor $\gamma=E/m_{0}c^{2} \gg 1$ the  equations of motion can be written as
\begin{equation}
	\frac{d\bmath{r}}{d(ct)} = \hat{\bmath{v}}
	\mathrm{~~~~~and~~~~~}
	\frac{d\bmath{\hat{v}}}{d(ct)} = \frac{q}{E} [\bmath{\hat{v}} \times \bmath{B}]~,
	\label{eq:motion}
\end{equation}
where $\bmath{r}$ stands for the radius-vector, $q = Ze$ is the electric charge, $\bmath{v}$ $(|\bmath{v}|\approx c)$ is the velocity (practically equal to the speed of light $c$), $\bmath{p} = \gamma m_{\mathrm{0}} \bmath{v}$ is the momentum and $\bmath{\hat{v}} = {\bmath{p}}/{|\bmath{p}|}$ is the unit velocity/momentum vector of a particle. The propagation of UHECR, governed by the regular GMF component, was calculated by solving the equations~(\ref{eq:motion}) numerically, using the 4th order Runge--Kutta method.

In our simulations we used the back-tracking method. Unlike the case of Centaurus~A, where clustering of UHECR events was observed in its closest vicinity \citep{2008APh....29..188P}, there are no such groups of events near Virgo~A. Hence it is difficult to assign some registered events as originating in Virgo~A, even if they are present indeed. Therefore we checked all the directions with a step of $0.5^{\circ}$. The motion of UHECR particles with different rigidities ${\cal R} = E/Ze$ was modelled. Within the Galaxy the simulated particles were initially moving under the influence of the regular GMF \citep{2012ApJ...757...14J} only, which was spatially limited by the sphere of radius $20 \mathrm{~kpc}$. Then the influence of the Galactic random fields \citep{2012ApJ...761L..11J} was accounted for as the overlapping of the scattering cone with the corresponding angle $\Theta$ (Eq. \ref{eq:scat-theta-Gal}) upon the outgoing direction. Additional scattering in the random EGMF was accounted for in a similar way (Eq. \ref{eq:theta-ran}, \ref{eq:scat-theta}).

\subsection{UHECRs from Virgo~A and the VC in the simplest void IGM model}
\label{subsec:CRinVoidModel}

\begin{figure}
\centering
\includegraphics[width=0.7\linewidth, clip]{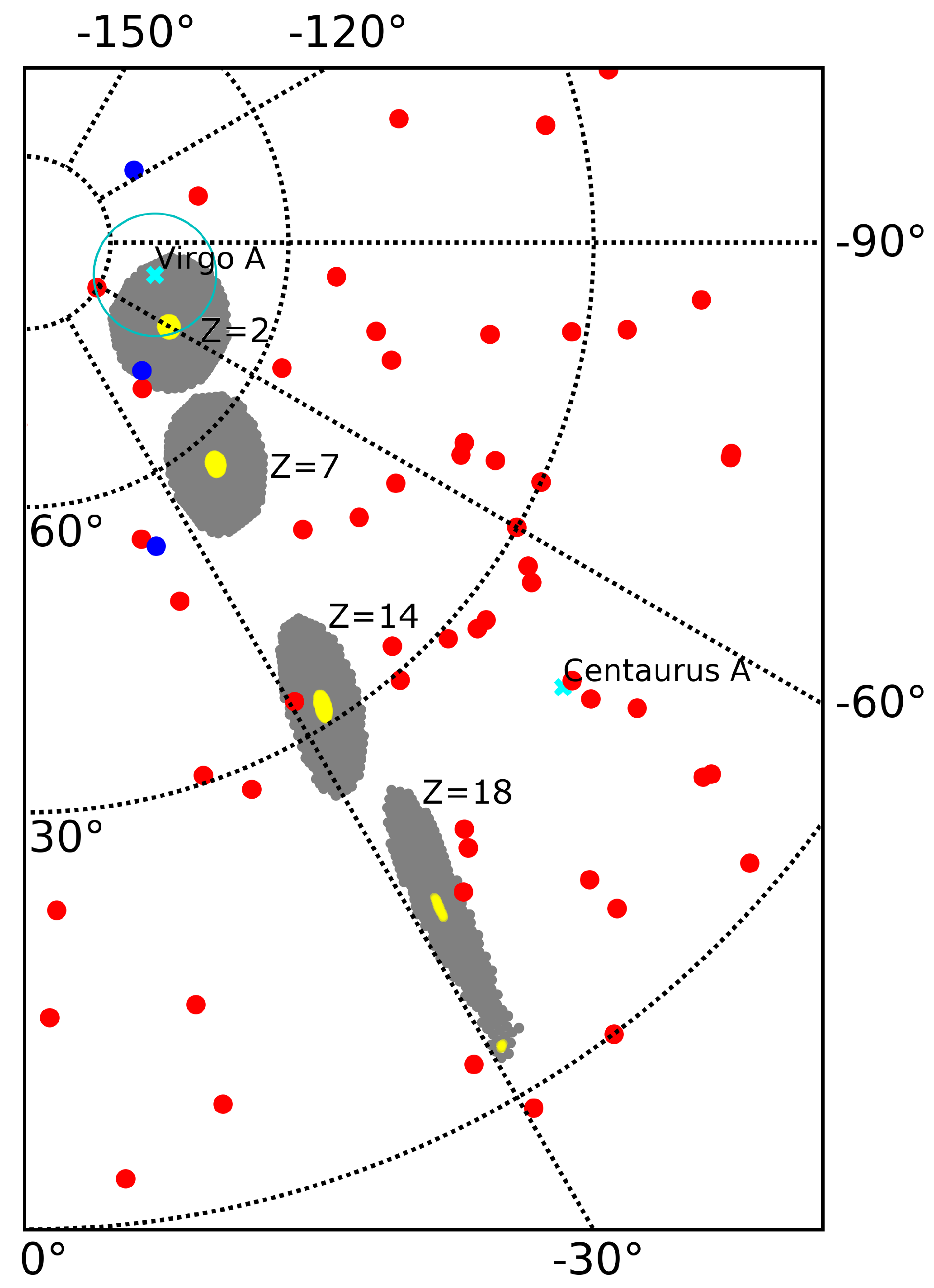}
\caption{Expected UHECR-images of Virgo~A (\textit{yellow}) and the VC (\textit{grey}) under the influence of the regular GMF only. Mono-energetic cases are presented for UHECRs with rigidity ${\cal R} = 10^{20}/Ze$ for: $Z = 2, 7, 14 \mathrm{~and~} 18$. Coloured circles denote the registered UHECR events: Auger (\textit{red}) and TA (\textit{blue}).}
\label{fig:img}
\end{figure}

\begin{figure*}
\centering
a)~\includegraphics[width=0.45\linewidth, clip]{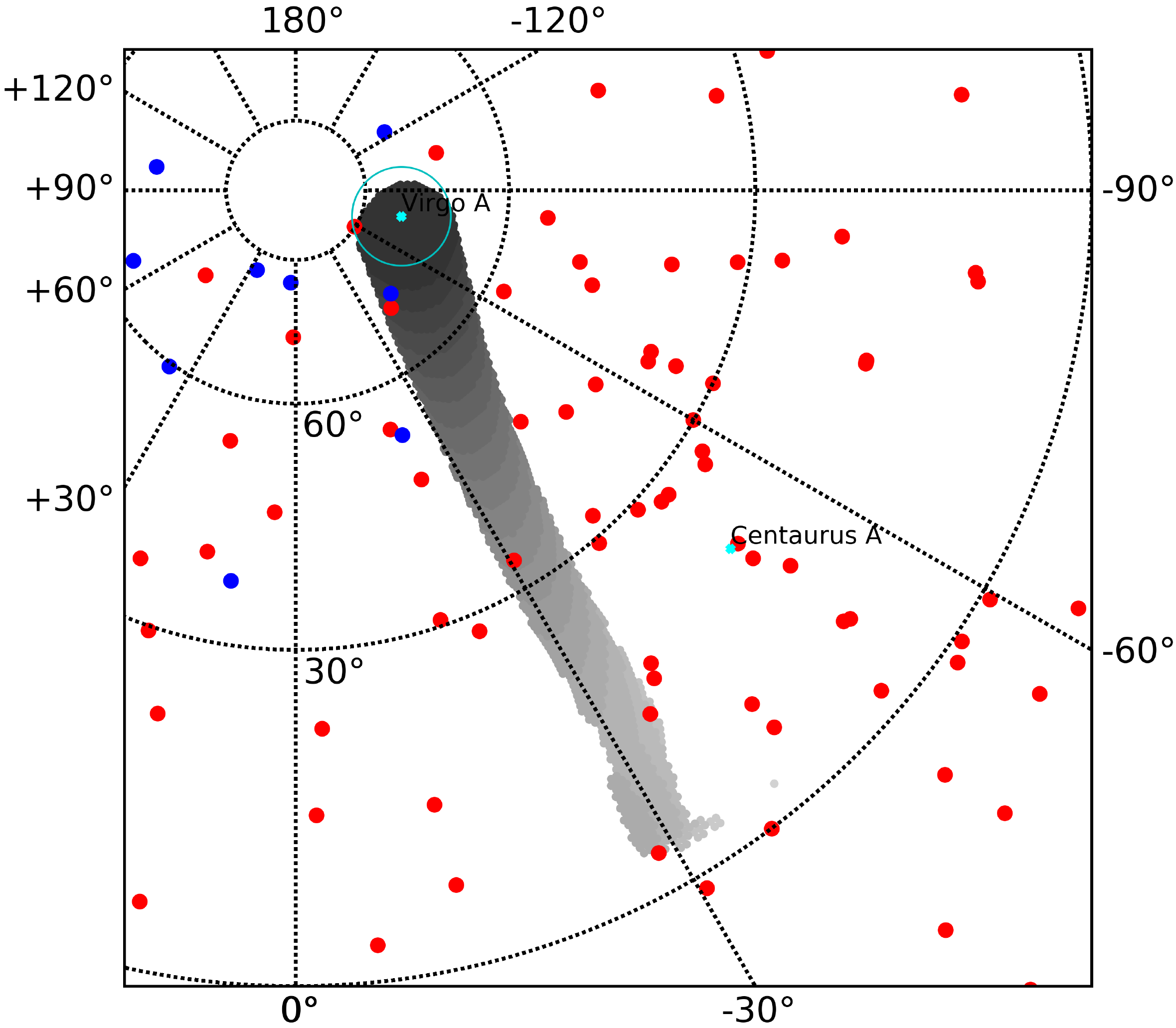}
~~~b)~\includegraphics[width=0.45\linewidth, clip]{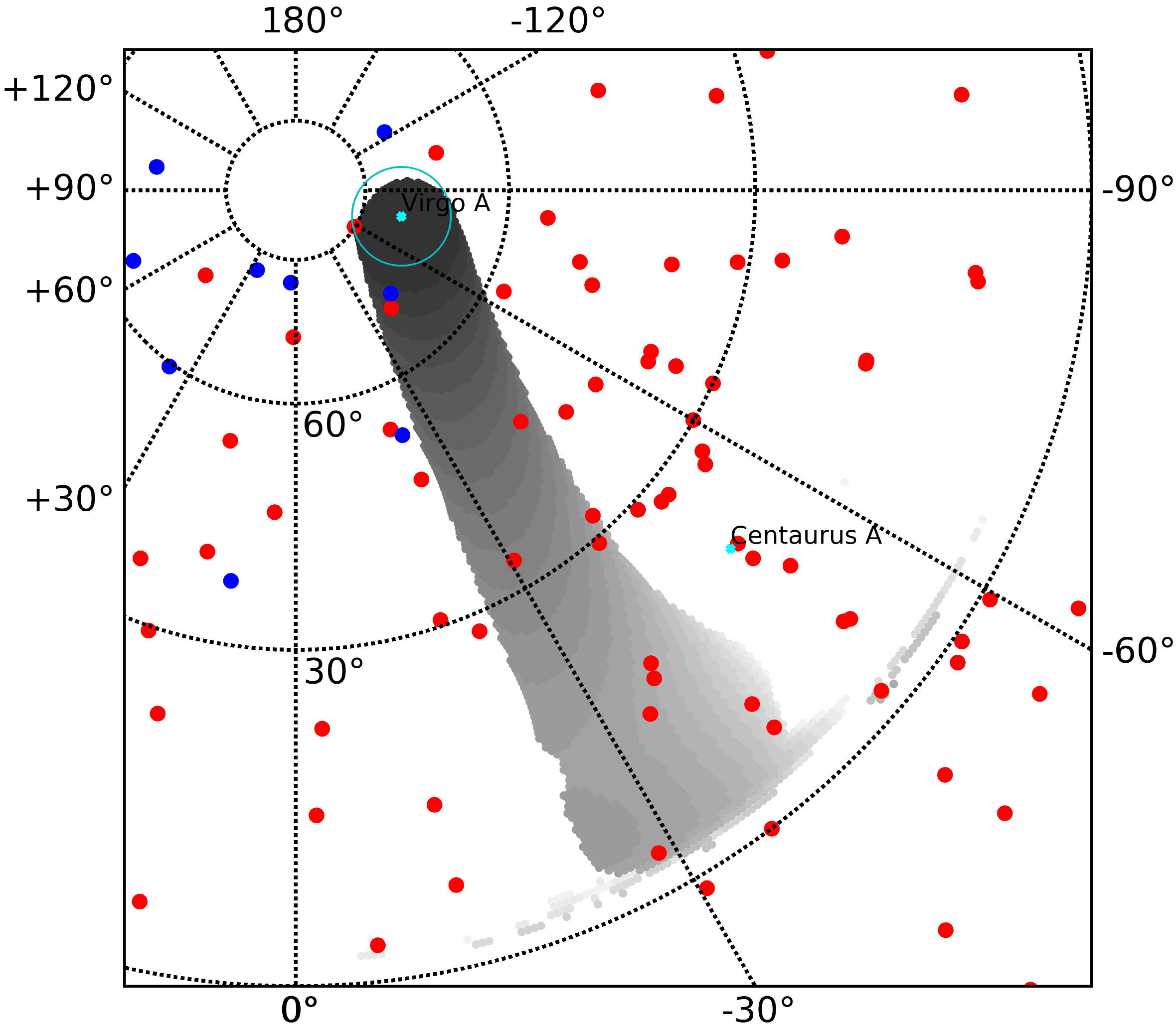}
c)~\includegraphics[width=0.45\linewidth, clip]{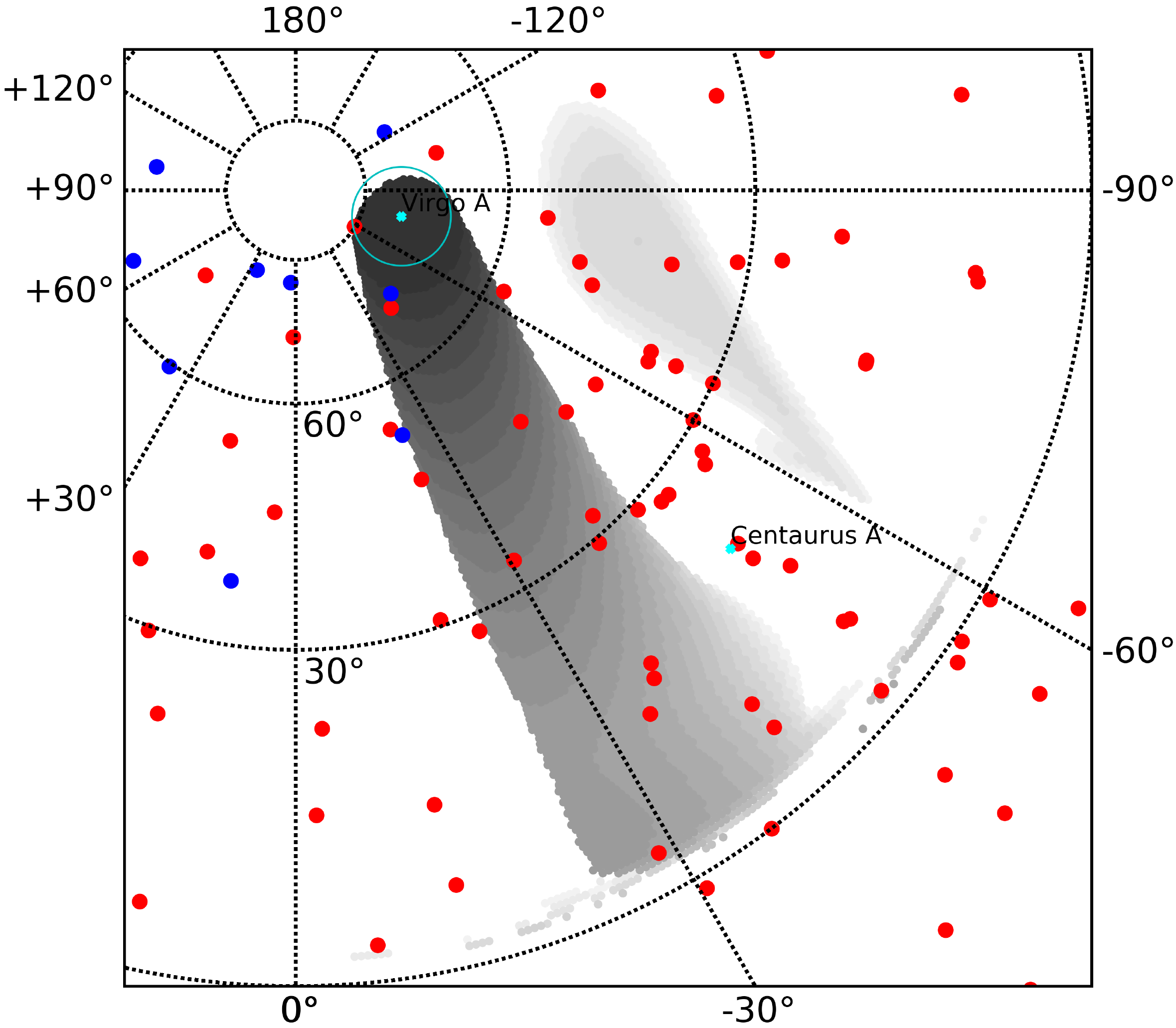}
~~~d)~\includegraphics[width=0.45\linewidth, clip]{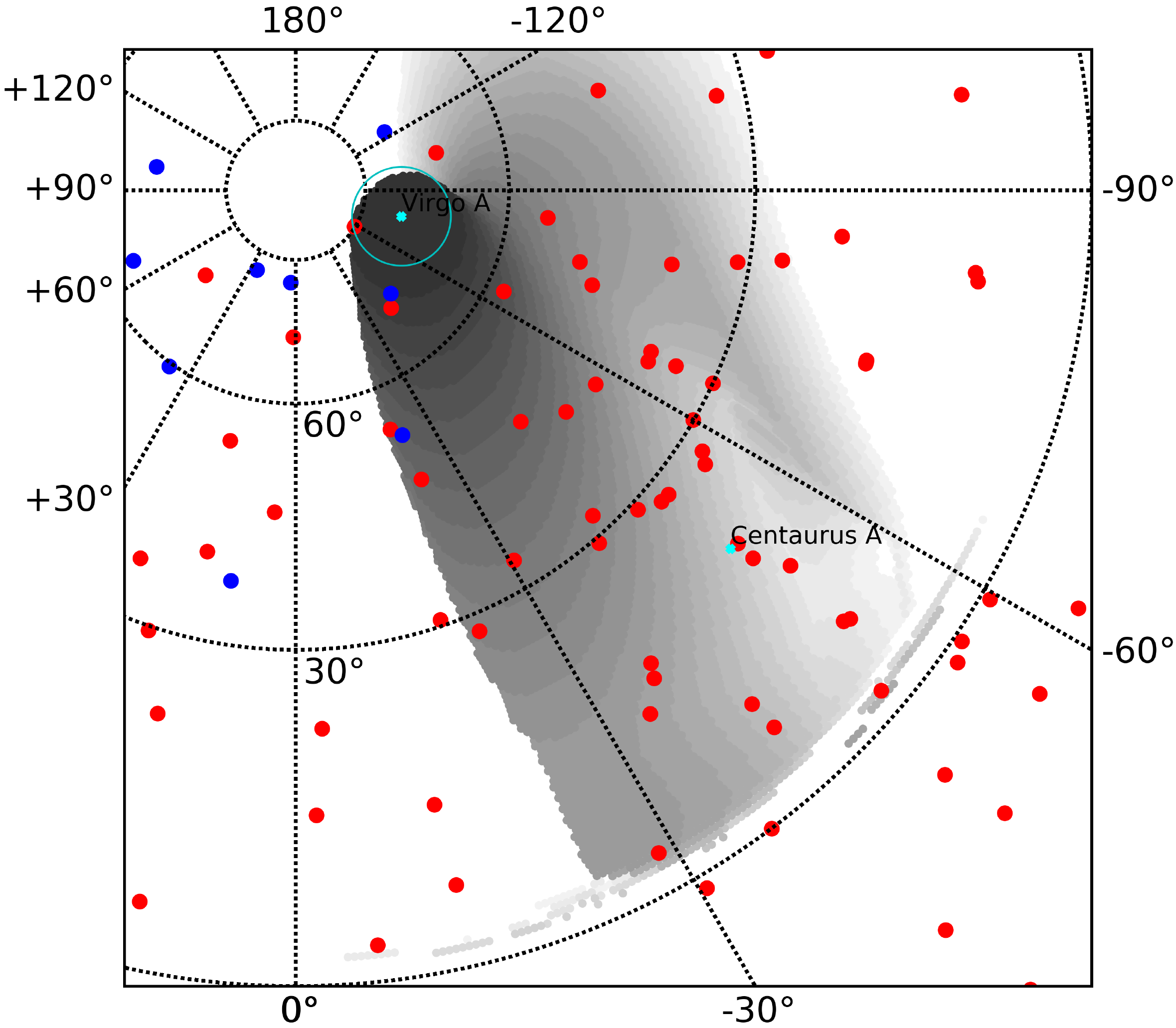}
\includegraphics[width=0.6\linewidth, clip]{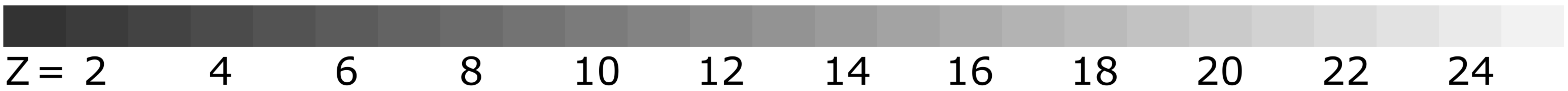}
\caption{Maps of the expected arrival regions for UHECRs with rigidity ${\cal R} = 10^{20}/Z$ V for $Z = 1~...~20~(a), Z = 1~...~25~(b-d)$, coming from 1 rad outskirt of Virgo~A and the VC. Gradient grey fields denote the calculated regions for events from the VC assuming its angular size $\sim 7^\circ$. Deflection and scattering in different MF components is accounted for: (\textit{a}) regular GMF only, (\textit{b}) regular + random GMF, (\textit{c}) regular + random GMF combined with the half-maximum EGMF, (\textit{d}) regular + random GMF combined with the upper limit EGMF. Level of grey corresponds to different $Z$: from 1 to 25 (see the gradient scale). Coloured circles denote the registered  Auger (\textit{red}) and TA (\textit{blue}) EHECR events.}
\label{fig:maps}
\end{figure*}

UHECRs accelerated in Virgo~A ($1^{\circ}$ angular radius) or in the VC volume, modelled by the sphere of radius 2~Mpc ($7^{\circ}$ angular radius), are injected into the intergalactic medium (IGM) and, being deflected by the IGMF and the GMF, can reach the  observer at Earth. Let us consider the simplest model of IGM between the VC and our Galaxy as a void with random magnetic field $B_{\mathrm{void}} = (5 - 10) \times 10^{-10} \mathrm{~G}$ of coherent length $l_{\mathrm{c, void}} = 1 \mathrm{~Mpc}$. 

In order to reveal the influence of the EGMF and the GMF on the arrival directions of EHECR from the VC, we analysed four different cases of the magnetic field configuration:
\begin{itemize}
\item regular (coherent) GMF only;
\item regular + random ($l_{\mathrm{c}} = 30 \mathrm{~pc}$) GMF;
\item regular + random GMF combined with the half-maximum EGMF ($B_{\mathrm{void}} = 5 \times 10^{-10} \mathrm{~G}$);
\item regular + random GMF combined with the maximum (upper limit) EGMF ($B_{\mathrm{void}} = 10^{-9} \mathrm{~G}$).
\end{itemize}

For each case we have carried out simulations and built the maps of expected arrival directions for UHECRs with rigidity ${\cal R} = E/Ze$ in the range $(2 - 100) \times 10^{18} \mathrm{~V}$. It was performed by choosing the energy of simulated particles $E = 10^{20} \mathrm{~eV}$ and varying their charge numbers from 1 to 50 in order to cover the expected rigidities of Auger ($E > 52 \mathrm{~EeV}$) and TA ($E > 57 \mathrm{~EeV}$) events with preferable for EHECRs charges from $Z = 1$ (protons) up to $Z = 26$ (Iron nuclei) $18.3 < \lg({\cal R}/V) < 20.0$.

The results obtained for UHECRs with total deflections up to $\sim 1$ rad under the influence of the regular (coherent) GMF (i.e., for UHECRs that are still related with their sources) are presented in Fig.~\ref{fig:img} in Galactic coordinates. 
We showed here only few cases of rigidity in the range ${\cal R} > 5 \times 10^{18} \mathrm{~V}$ to illustrate the formation of Virgo~A and the VC images by Auger and TA representative set of nuclei (H, He, N, Si, Fe) with energy $E=10^{20}$ eV. 
As one can see, at higher rigidities the image only shifts a little from the source position, being accompanied with a minor distortion. With the decrease of ${\cal R}$ the distortion grows essentially, resulting in the appearance of a multi-image at ${\cal R} \approx 6 \times 10^{18} \mathrm{~V}$. 
With further rigidity reduce the image starts to decrease and disappears at ${\cal R} \approx 5 \times 10^{18} \mathrm{~V}$. It means that the events with $E = (5 - 10) \times 10^{19} \mathrm{~eV}$ can be observed at such conditions only till $Z \approx (10 - 20)$.

Such a result is caused by lensing of UHECR trajectories in the  regular GMF, corresponding deflection, spreading, multiple imaging, and magnification or demagnification of the UHECR flux depending on arrival directions, and agrees with the results of detailed calculations of UHECR deflection in the GMF, presented in \cite{2017arXiv171102730F} (see Fig. 12 and Table 4 there for the Virgo~A case). 
As  shown in \citep{2017arXiv171102730F}, strong demagnification of the Virgo~A flux by the regular GMF takes place in  ${\cal R} \sim (2.5 - 5) \times 10^{18} \mathrm{~V}$ range.

The influence of the random field components is illustrated in Figure~\ref{fig:maps}. 
The case of the regular GMF only is presented for comparison in Fig.~\ref{fig:maps}a. 
It is built of the overlapped images of the VC obtained for rigidities in the range ${\cal R} = (5 - 100) \times 10^{18} \mathrm{~V}$ ($Z = (1 - 20)$ for $E = 10^{20} \mathrm{~eV}$), shown with grey areas gradually decreasing in intensity as $Z$ increases.

Accounting for the random GMF in addition to the regular one modifies regular field case (Fig.~\ref{fig:maps}a) due to CR scattering (Fig.~\ref{fig:maps}b).
At first, it eliminates the  demagnification effect and makes the ${\cal R} \leq 5 \times 10^{18} \mathrm{V}$ CRs available to reach the Earth as well. 
Here (Fig.~\ref{fig:maps}b-d) we present the results for some wider range of rigidities ${\cal R} = (4 - 100) \times 10^{18} \mathrm{~V}$ ($Z = (1 - 25)$ for $E = 10^{20} \mathrm{~eV}$). It also leads to the significant widening of the "expectation regions", especially for smaller ${\cal R}$. 
As one can see, if we account for the random GMF only, the expected UHECR image remains practically single, like in the case of the regular GMF only. 
In the following such type of an image is referred to as "main track". 
It is worth to note, that this result, obtained for $l_{\mathrm{c}} = 30 \mathrm{~kpc}$ case, is very sensitive to $l_{\mathrm{c}}$ value and for $l_{\mathrm{c}} = 100 \mathrm{~kpc}$ multiple images arrive even at ${\cal R} = 10^{19} \mathrm{~V}$ \citep{2017arXiv171102730F}.

Meantime, it was found that the influence of the random EGMF in the weak field case ($B_{\mathrm{void}} \sim 10^{-10} \mathrm{~G}$) can be neglected compared to the random GMF ($\Theta_{\mathrm{rms}}({\cal R})\sim 1^{\circ} ({\cal R}/10^{19} \mathrm{~V})^{-1}$, see Eq. \ref{eq:theta-ran}). 
Increase of the EGMF to $B_{\mathrm{void}} \sim 5 \times 10^{-10} \mathrm{~G}$ results not only in the widening of the main track (since $\Theta_{\mathrm{rms}} \propto B_{\mathrm{void}}$), but also in the appearance of a multi-image (light-grey spot at the left-hand side of the main track in Fig.~\ref{fig:maps}c). 
In the following this type of an image is referred to as "side spot".
Only increase of the EGMF to the upper-limit amplitude ($B_{\mathrm{eg}} = 10^{-9} \mathrm{~G}$) makes its impact quite significant and comparable to the random GMF (Fig.~\ref{fig:maps}d).
It results in the considerable widening of the side spot and its merging with the main track (Fig.~\ref{fig:maps}d), while the last one changes not that much. 

We can compare now Virgo~A and the VC ${\cal R}$-dependent images with the observed arrival directions of the TA and Auger EHECRs (Fig.~\ref{fig:maps}). The main results of this comparison  were noted earlier \citep{2015arXiv150909033S,insert1}: the comet-like image of Virgo~A/VC is stretched with the decreasing of CR rigidity to the lower Galactic latitudes with no evident enhancement of events from this region detected. 
Once more, there are no p or He $ E> 52$  EeV EHECRs (whose deflections are minor)  from Virgo~A and the VC. 
CNO - nuclei with  $\lg ({\cal R}/V)>18.8$  and Fe nuclei with  $\lg({\cal R}/V) > 18.3$ for our dataset, if any,  arrive from the Centaurus~A region.
For supported by Auger He ($\lg({\cal R}/V) > 19.4$) -- Si ($\lg({\cal R}/V) > 18.6$ chemical composition estimation \citep{2018arXiv180101854T} no event enhancements can be related with Virgo~A/VC sources. 
It is worth to note that one of the reasons for the absence of enhanced  EHECR flux from Virgo~A/VC, especially, of heavy nuclei flux, is the position of the VC in a demagnification zone, where the regular GMF acts as a magnetic lens and for small ${\cal R}$ can considerably reduce the EHECR flux \cite{2017arXiv171102730F}.

\section{UHECRs from Virgo~A and the VC in void and filament model}
\label{sec:filamentmodel}

\begin{figure*}
\centering
a)~\includegraphics[width=0.4\linewidth,clip]{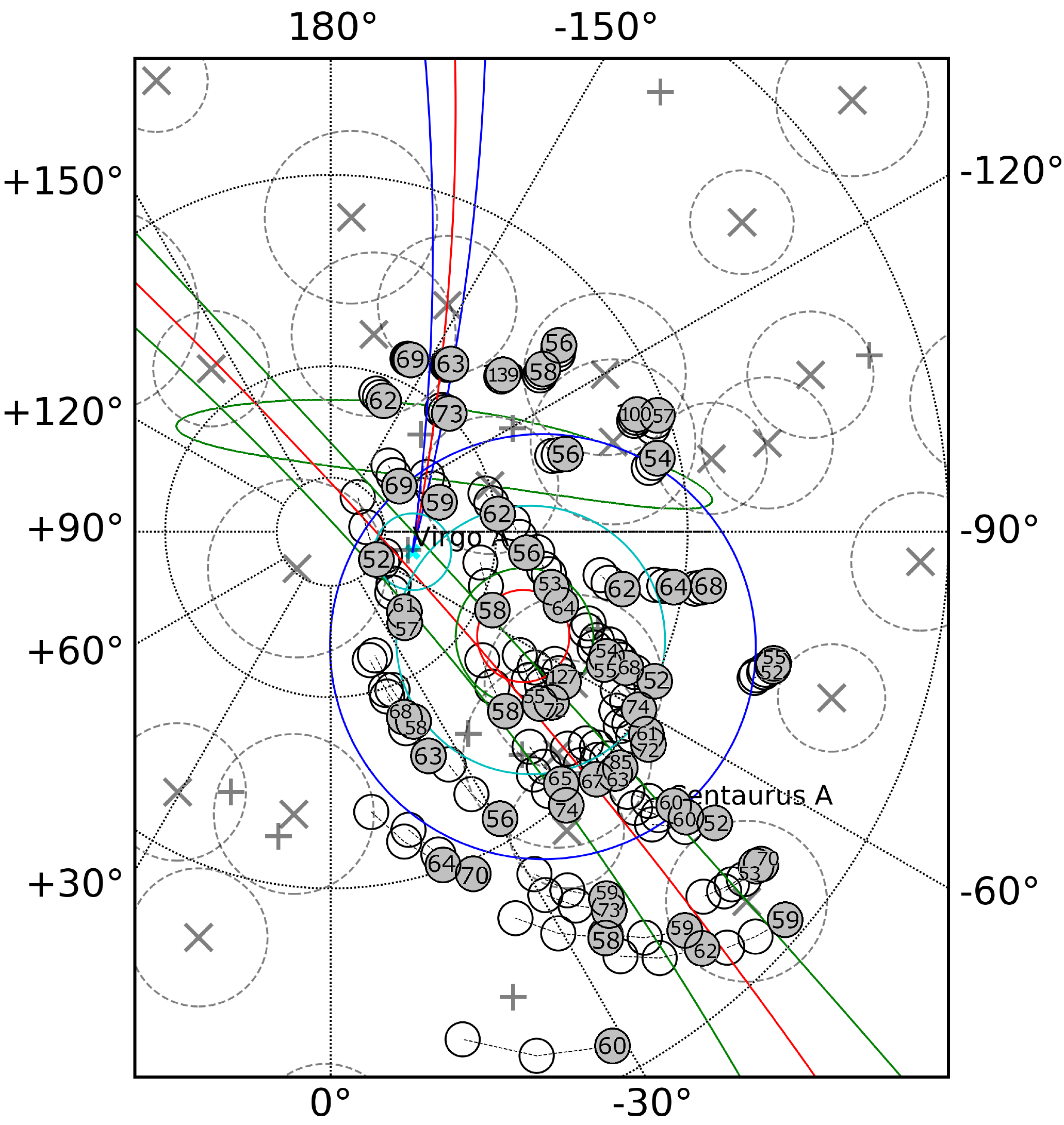}
~~~~b)~\includegraphics[width=0.4\linewidth,clip]{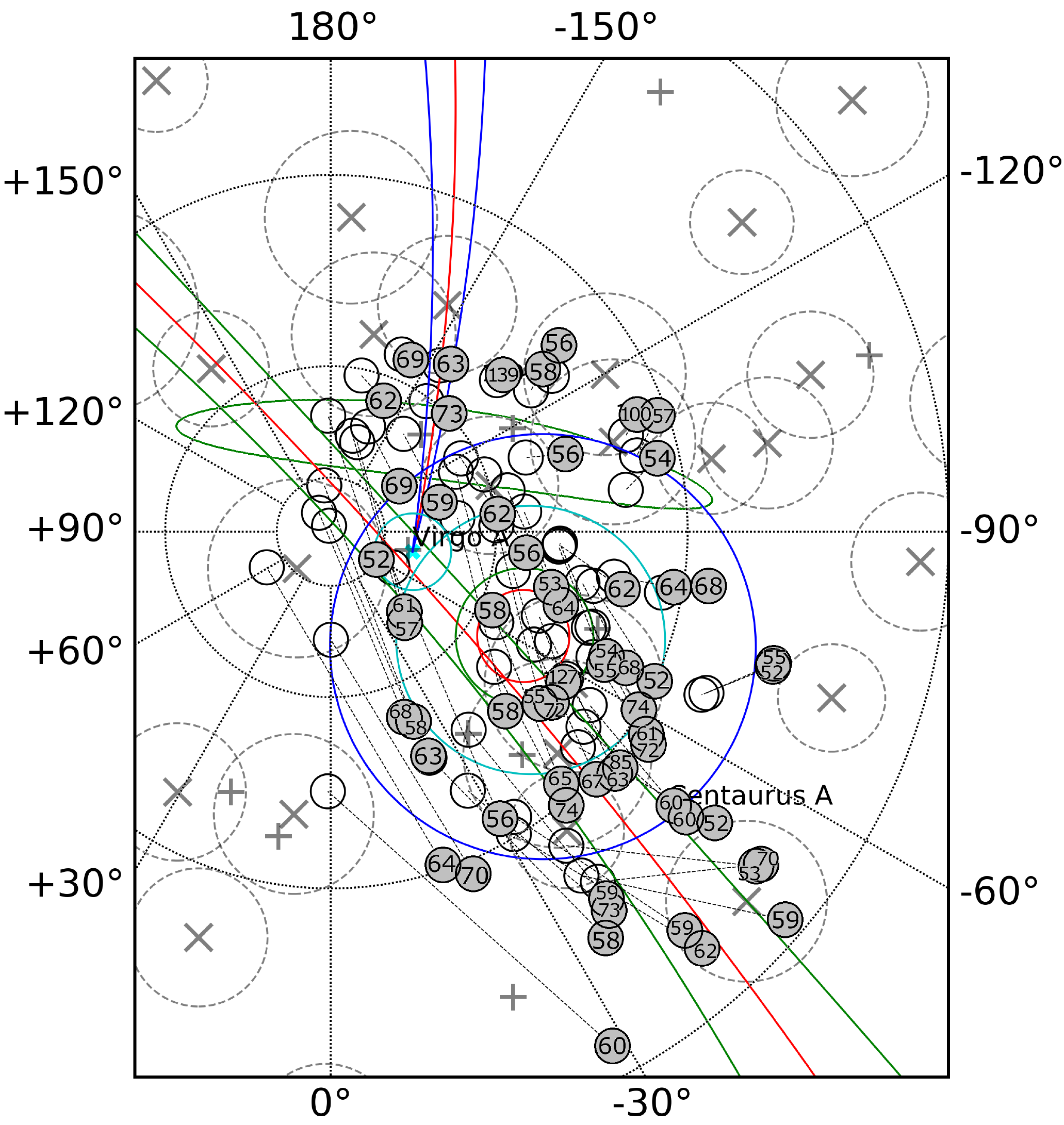}
c)\includegraphics[width=0.4\linewidth,clip]{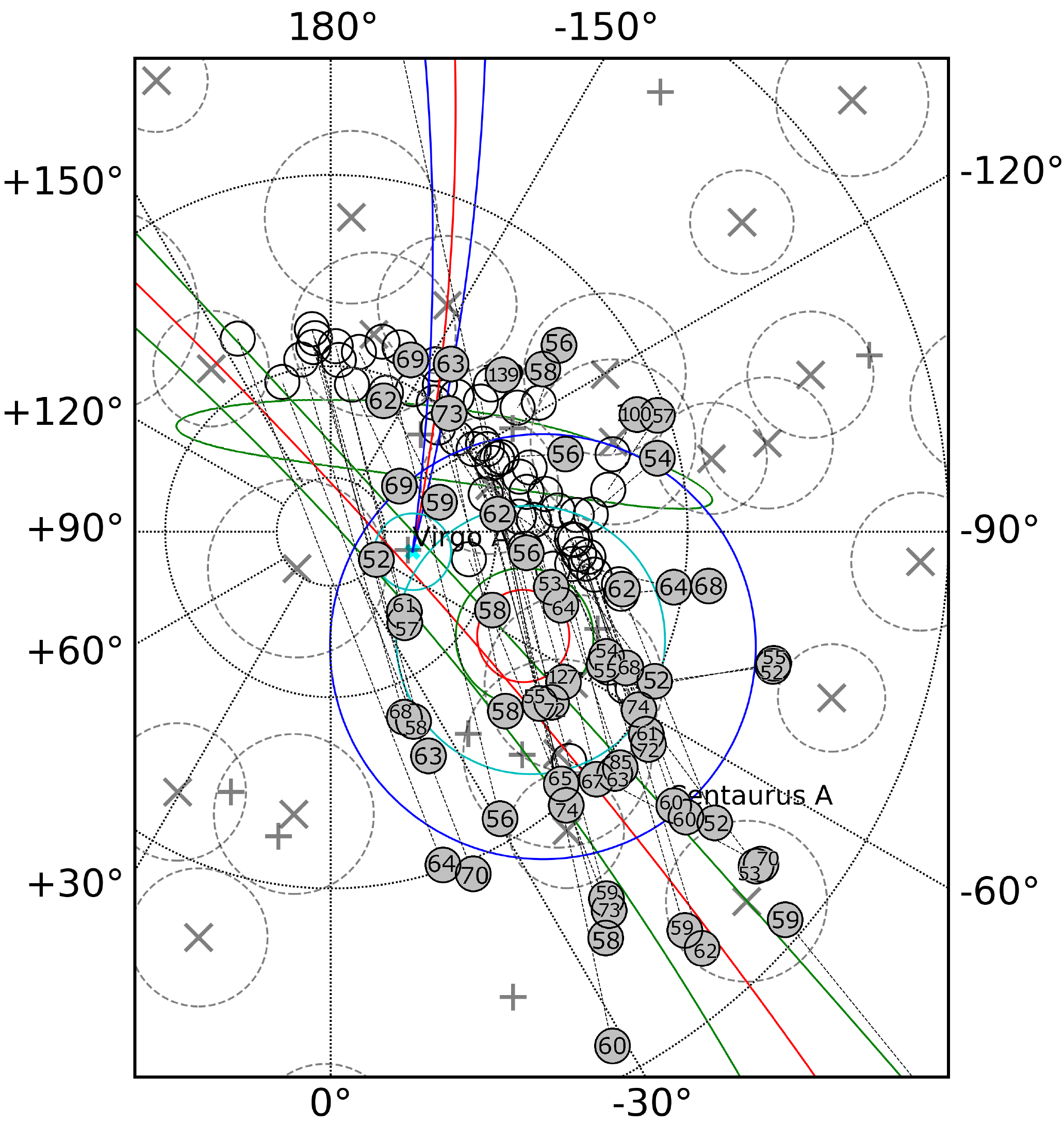}
~~~~d)~\includegraphics[width=0.4\linewidth,clip]{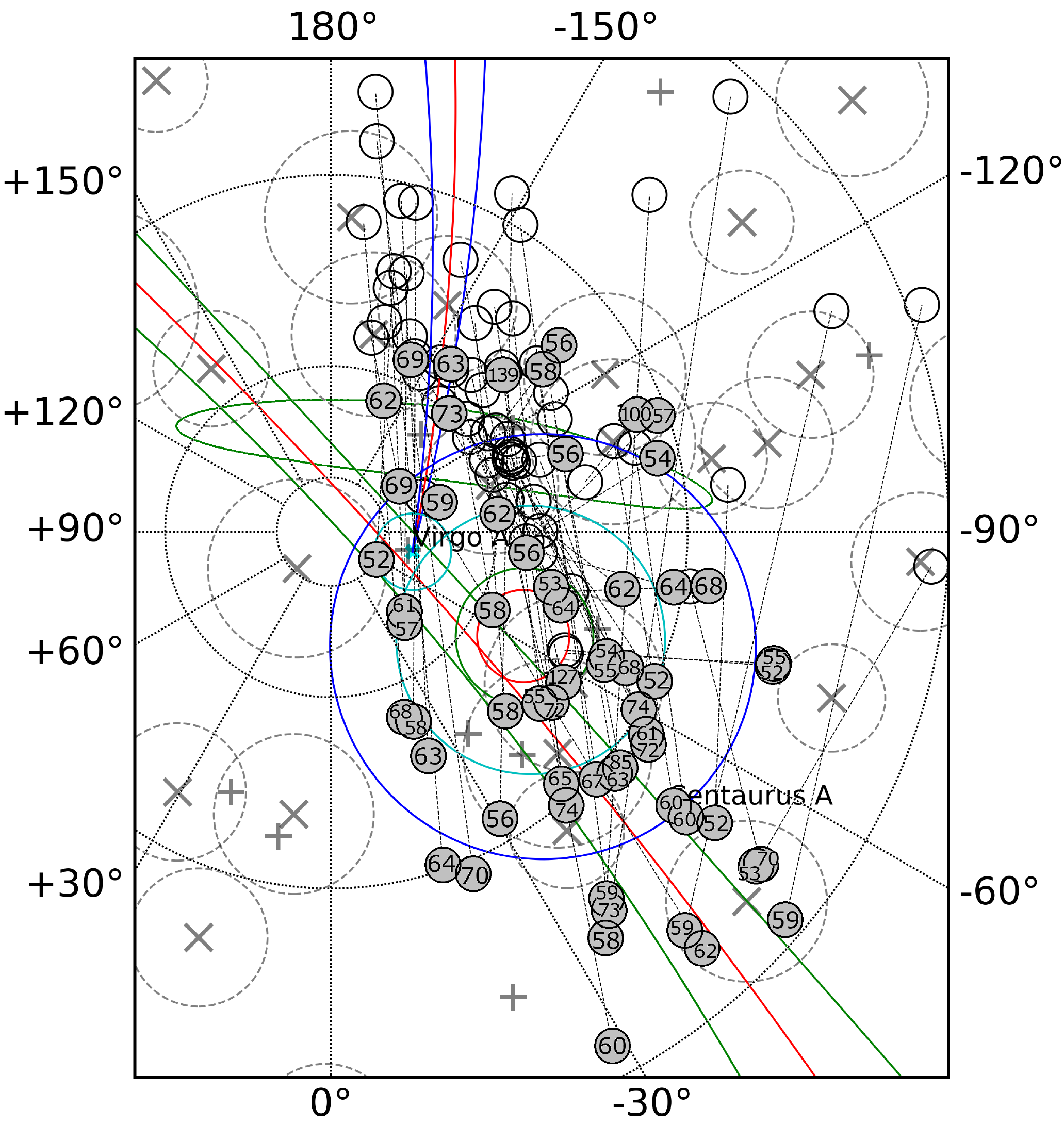}
\caption{Maps of expected UHECR arrival directions at the Galaxy boundary from the LF region. Notations as in Fig.~\ref{fig:vs}, but Auger and TA events are indicated by the grey circles with observed energy (in EeV). Event-by-event calculated arrival directions at the Galactic boundary are shown for the case of H and He (\textit{a}), N (\textit{b}), Si (\textit{c}), and Fe (\textit{d}) nuclei.}
\label{fig:lf}
\end{figure*}

As one can see from Fig.~\ref{fig:maps} and Eq.~\ref{eq:theta-ran}, the typical deflection of EHECRs in the IGMF from the sources in the LU ($L<100 \mathrm{~Mpc}$) is of the order of a few degrees even for heavy nuclei $\Theta \leq 0.5^{\circ} Z$ for $B\leq 10^{-10} \mathrm{~G}$. 
Meantime, considerably larger deflections are  accompanied by EHECR crossing the elements of the LSS - sheets, filaments, nodes (clusters). 
While there is a small probability of such a crossing in the VSC and even in the LU, the arrival directions of EHECRs, which indicate the centres of their last scatterings  \citep{2008PhRvD..77l3003K}, should correlate with close elements of the LSS. Within the VSC the Local Filament may act as such a scattering centre (Fig.~\ref{fig:lf}).

In order to investigate the possible role of the LF in EHECR deflection we have calculated the arrival directions of the EHECR from the LF region at the entrance into the Galaxy (Fig.~\ref{fig:lf}), using an event-by-event approach. 
The common feature of the regular GMF in this region is the decrease of Galactic latitudes of EHECR arrival directions together with additional increase of their Galactic longitudes, especially for large $Z$ (small ${\cal R}$). 
In cases of light (p, He, Fig.~\ref{fig:lf}a) and intermediate (N, Fig.~\ref{fig:lf}b) EHECR composition there is a clear trend of extragalactic arrival directions to the LF position.
In a case of Si nuclei (Fig.~\ref{fig:lf}c) 9 events penetrate Galaxy from the Hot Spot position, the others create the elongated hot spot around $l \sim 270^{\circ}$, $b \sim 50^{\circ}$, i.e., in the LF and the Local Sheet directions again.
For Fe nuclei (Fig.~\ref{fig:lf}d) extragalactic arrival directions show similar behaviour: contribute to the Hot Spot  region and an arc-shaped structure  around $l \sim 230^{\circ}$, $b \sim 60^{\circ}$, partially overlapping with the LF and the LS regions.

It is worth to note that these elongated spot-like patterns of EHECR extragalactic arrival directions are caused by the structure of the regular GMF component in the LF region, namely, arriving of elongated zone $l \sim 150^{\circ} - 300^{\circ}$, $b \sim 30^{\circ} - 60^{\circ}$ of strong magnification of the CR flux for $\lg({\cal R}/V) \leq 19.0$ \citep{2017arXiv171102730F}. 
EHECR arrival directions from the sources in this region, due to deflection in regular and random components of the GMF, become more and more isotropic with the decrease of their rigidity ${\cal R}$, providing observed arrival directions of low-${\cal R}$ EHECRs from the physically demagnification regions.  

Taken together, the obtained results substantiate a notable influence of the LF and the Local Sheet on the observed EHECR flux via deflection of EHECR trajectory from the distant sources and/or contribution from EHECR sources inside them. The presence of neutrino events from the LF directions also agrees with this interpretation.  

\section{EHECRs from Virgo~A jet in void and jet model}
\label{sec:jetmodel}

\begin{figure*}
\centering
a)~\includegraphics[width=0.4\linewidth,clip]{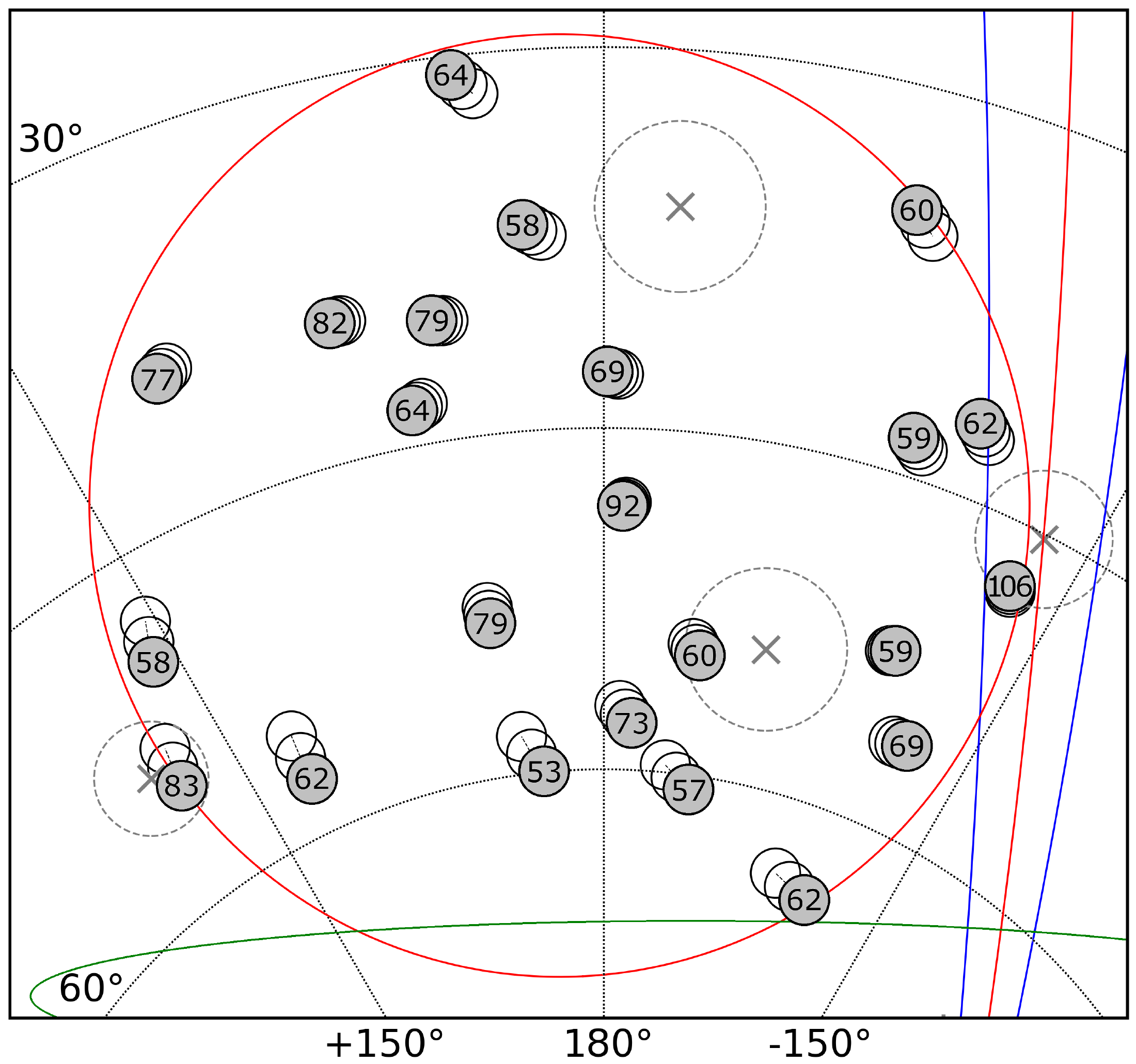}
~~~~b)~\includegraphics[width=0.4\linewidth,clip]{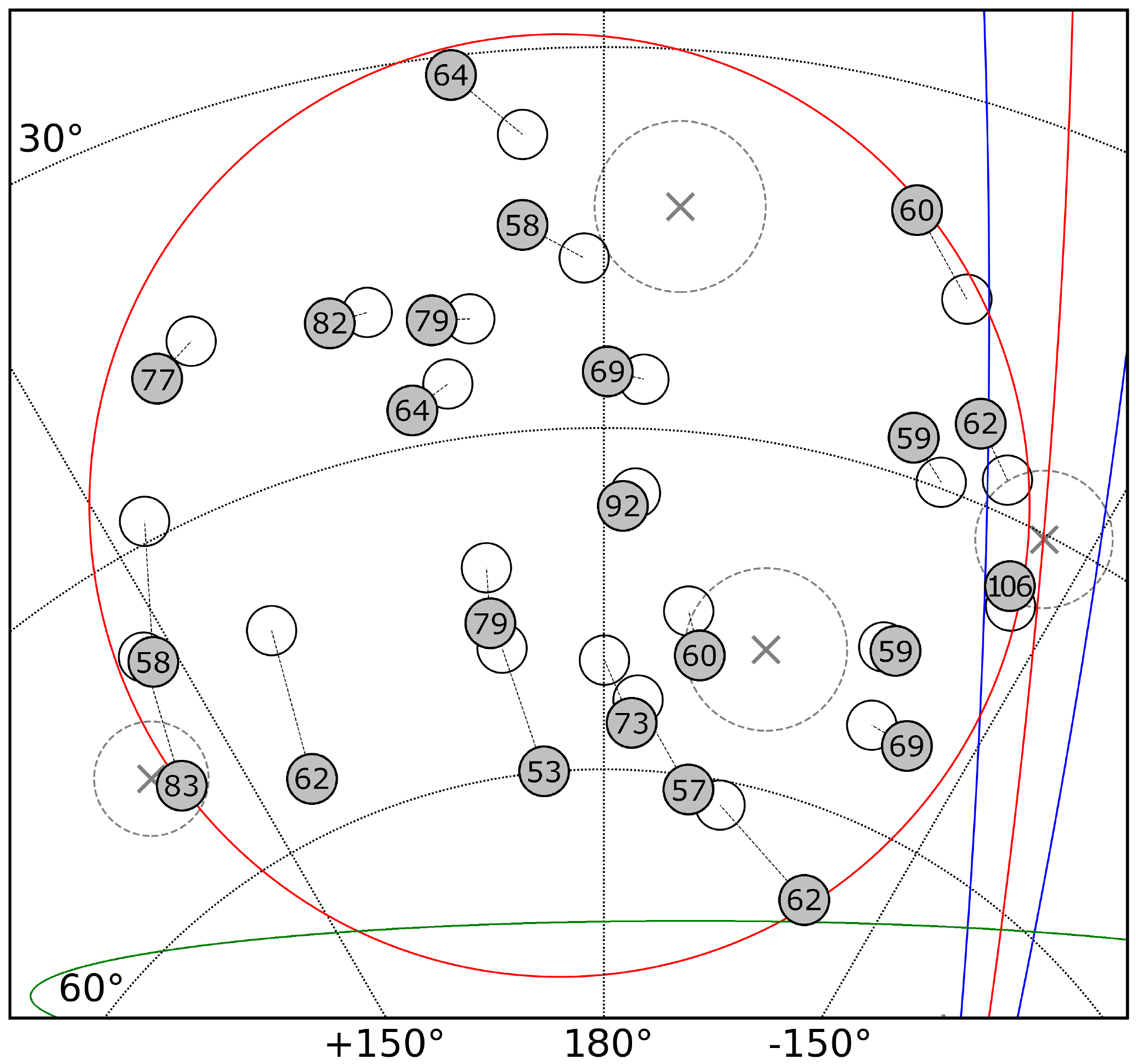}
c)~\includegraphics[width=0.4\linewidth,clip]{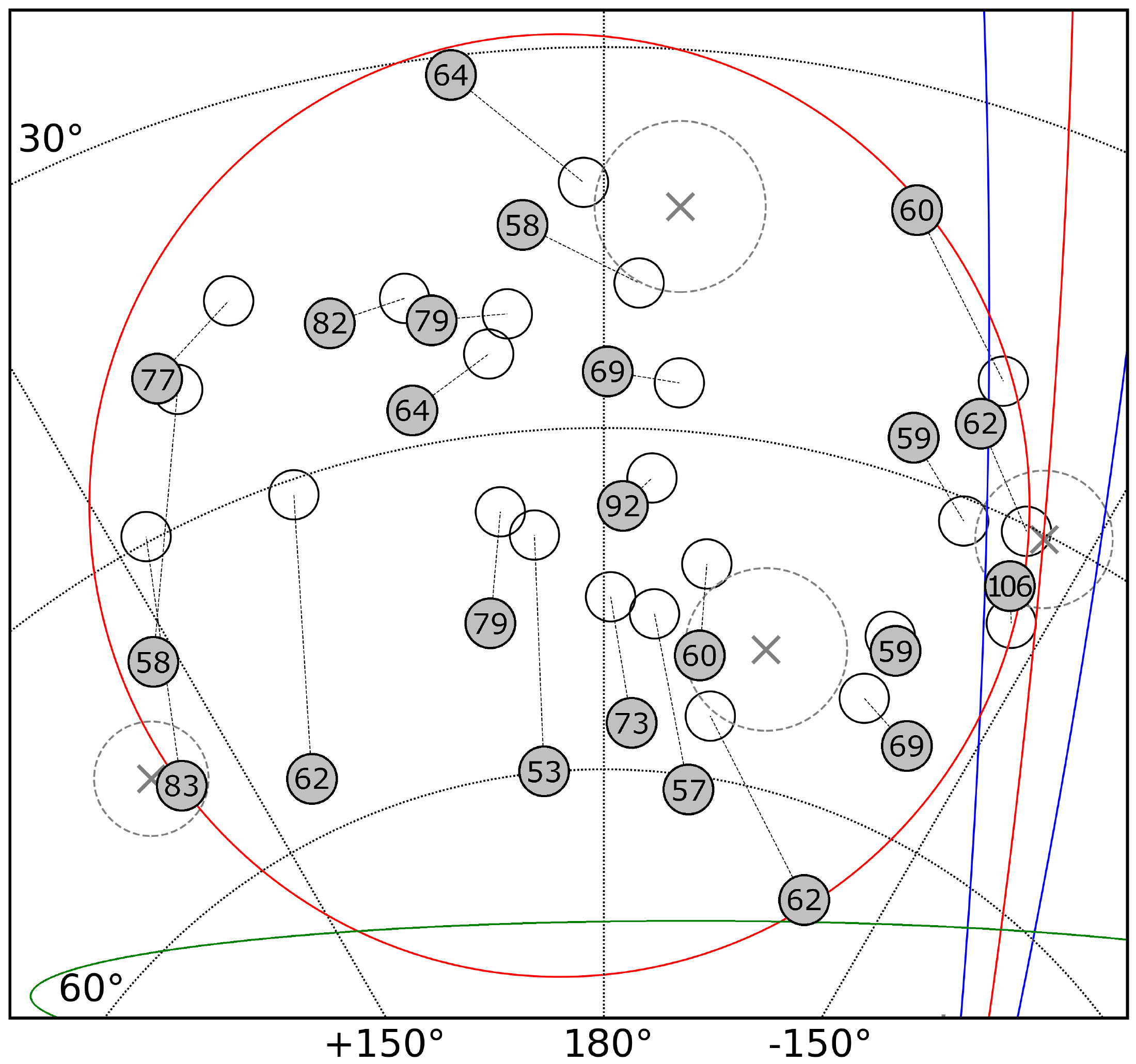}
~~~~d)~\includegraphics[width=0.4\linewidth,clip]{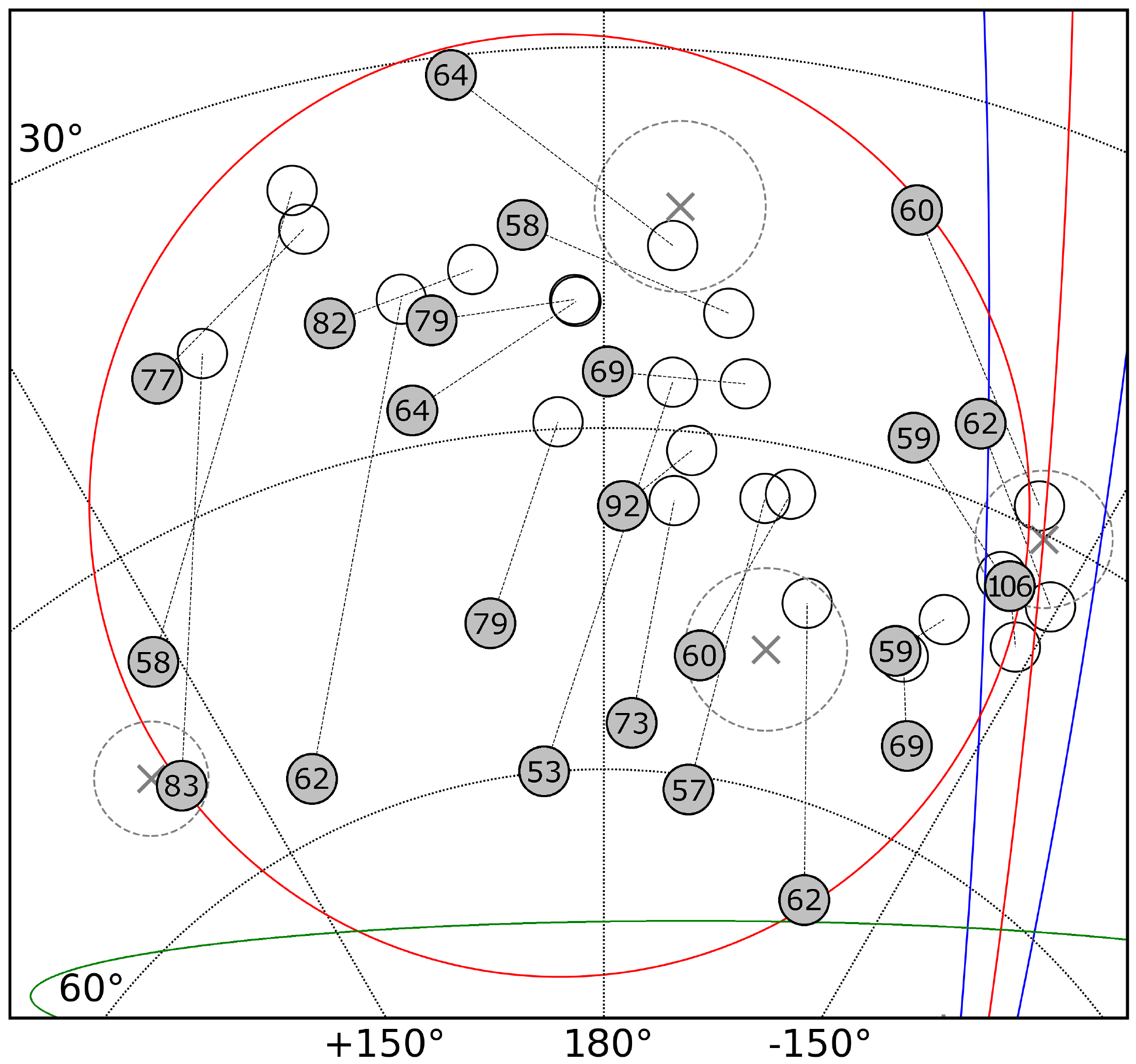}
\caption{Maps of expected UHECR arrival directions at the Galaxy boundary from the Hot Spot region. Notations as in Fig.~\ref{fig:lf}. Event-by-event calculated arrival directions at the Galactic boundary are shown for the case of H and He (\textit{a}), N (\textit{b}), Si (\textit{c}), and Fe (\textit{d}) nuclei.}
\label{fig:hs}
\end{figure*}

Two regions of EHECR enhancement -- the Hot Spot and the described above spot-like elongated zone -- create a belt-like structure $ \sim 90^{\circ}$ long and $ \sim 20^{\circ}$ wide, transversal to the Virgo~A jet sky projection (Fig.~\ref{fig:lf}c-d). The Hot Spot is usually considered as the product of a starburst galaxy M82 \citep[see, i.e.][]{2016PhRvD..93d3011H}, but correction of EHECR arrival directions from the Hot Spot taking into account the influence of the regular GMF, especially for intermediate and heavy nuclei (Fig.~\ref{fig:hs}), weakens the Hot Spot-M82 correlation and strengthens the Hot Spot-jet correlation.   

AGN jets are well-known potential sources of UHECRs, especially in the case of giant AGN flares \citep[see, i.e.][]{2009ApJ...693..329F,2012A&A...541A..31N,2015ApJ...803...15P,2017ApJS..232....7F}. Virgo~A has experienced such a $\sim$ 2 Myr flare $\sim$ 10 -- 12 Myr ago, having allocated $5 \times  10^{57}$ erg of energy \citep{2017ApJ...844..122F}. If UHECRs were accelerated in an ultrarelativistic jet of high \ok{bulk} Lorentz-factor $\Gamma \gg 1$, their initial trajectory will be collimated within a narrow cone $\Theta_{\mathrm{jet}} \simeq \Gamma^{-1}$ and, with time, this transient jet will widen due to small-angle scattering on IGMF
inhomogeneities. For fiducial EHECR energy $E = 6 \times 10^{19} \mathrm{~eV}$ and coherence length $l_\mathrm{c} = 1 \mathrm{~Mpc}$ the mean deflection of the momentum (velocity) unit vector in the EHECR jet near the Local Group is expected to be $\Theta_{\mathrm{rms}} \sim 1.7^{\circ} ZB_{-9}$ and mean time delay $\tau_{\mathrm{rms}} \simeq 1.2 \times 10^4 Z^2 B_{-9}^2$ yr (Eq. \ref{eq:theta-ran}, \ref{eq:tau-ran}). Meantime, mean space widening (angle of rms displacement) of UHECR will be only 0.56 of angle of \textit{rms} velocity $\eta_{\mathrm{rms}} = 1/\sqrt{3} \times \Theta_{\mathrm{rms}}$ \citep{2002JHEP...03..045H}. Therefore, the condition of the most effective EHECR detection is their arrival direction at the Galaxy boundary $\Theta_{\mathrm{rms}} - \eta_{\mathrm{rms}} \simeq \theta_{\mathrm{jet}} \simeq 20^{\circ}$ or $ZB_{-9} \simeq 28$ together with mean time delay $\tau_{\mathrm{rms}} \simeq 9 {\mathrm{~Myr}}$. Therefore, we estimate the "jet induced Hot Superspot (JIHS)" as an ellipse with angular radius $15^{\circ} - 25^{\circ}$ along the jet directions and $\pm 45^{\circ}$ width transversal to the jet direction. This JIHS is visible in Figure~\ref{fig:lf} for the LF region and in Figure~\ref{fig:hs} for the Hot Spot region. As one follows from $ZB_{-9} \simeq 28$ restriction, we cannot expect to detect protons or He nuclei due to IGMF limitation, but for N ($Z = 7$), Si ($Z = 14$) and Fe ($Z = 26$) nuclei the necessary IGMF value is in the reasonable range $B_{-9} \simeq 4,~2,~1$ correspondingly. 

It is worth to note that we do not look for the best fit of predicted JIHS position  with extragalactic EHECR enhancement because both of them are strongly model-dependent.

\section{Discussion and Conclusions}
\label{sec:DisCon}

Galactic and intergalactic magnetic fields are believed to provide the observed isotropy of the UHECR flux. Therefore, the promising way to search for UHECR astrophysical sources is to analyse correlations of highest energy events with the nearest potential sources. 

\bh{
Among AGN such sources are, particularly, powerful radiogalaxies in Local Universe: Centaurus~A, Virgo~A, Fornax~A, Centaurus~B, IC310. All these sources are observable by Auger (IC310 only in inclined mode), and only Virgo A and IC310 by TA. As we mentioned earlier, no robust signs of UHECR excess from these sources were found in the TA and Auger data. However, some hope was given by recent Auger results \citep{2018ApJ...853L..29A}, in which two "hot spots" formed by $E>60$ EeV EHECR arrival directions can be preliminarily explained as contribution from Centaurus~A and Fornax~A \citep{2018MNRAS.479L..76M}
}

Here we revise this result for Virgo A/VC taking into account deflections of EHECR from the TA ($E > 57 \mathrm{~EeV}$) and Auger ($E > 52 \mathrm{~EeV}$) data sets in Galactic and extragalactic (VSC) magnetic fields. Even for such EHECRs regular and random GMF components significantly affect EHECR trajectories, progressively (with the decrease of EHECR rigidity) lowering galactic latitudes of their arrival directions (via the regular component) and spread displaced source's images (via the random component) (Fig.~\ref{fig:maps}a,b), but demagnification effect of the regular GMF in Virgo~A direction \citep{2017arXiv171102730F} partly compensates this spread, especially for small rigidities. Random void-type EGMF additionally spreads the source's image (Fig.~\ref{fig:maps}c,d).

In Fig.~\ref{fig:maps} we do not see any enhancement of arrival directions of the TA and Auger EHECRs along the train of Virgo A/VC images, i.e., we cannot confirm Virgo A/VC as a permanent isotropic source of EHECRs, surrounded by a typical intergalactic void. Therefore we consider a more realistic structure of the VSC taking into account the LF, whose enhanced magnetic field acts as the last scattering centre \citep{2008PhRvD..77l3003K}. We have calculated on event-by-event basis the extragalactic arrival directions of the detected EHECRs from the LF region. It is shown that in cases of light (p, He) and intermediate (N) nuclei there is a clear trend of extragalactic arrival directions to the LF position (Fig.~\ref{fig:lf}). It means that the LF can be an effective scatterer of EHECRs and/or contain the sources of EHECRs (AGNs, starburst galaxies, etc.). Additional independent argument supporting this assumption is some enhancement of neutrino events from the LF sky position (Fig.~\ref{fig:lf}).
   
Somewhat unexpected result of our event-by-event reconstruction of extragalactic arrival directions of detected EHECRs from the Virgo A/VC neighbourhood was an identification of "Hot Superspot", partially overlapping with the TA Hot/Cold Spot, -- a belt-like cluster of Si-Fe EHECRs $ \sim 90^{\circ}$ long and  $ \sim 20^{\circ}$  wide, transversal to the Virgo A jet sky projection (Fig.~\ref{fig:lf}~-~\ref{fig:hs}). We show that the Hot Superspot may be naturally explained as a jet induced transient event, created by a pancake-like beam of EHECRs, accelerated and collimated during the Virgo~A jet flare $\sim$10--12 Myr ago and spread en route due to small-angle scattering on IGMF inhomogeneities.

\section*{Acknowledgements}
\bh{The authors thank the anonymous referee for constructive and helpful comments.}
Work of O.\,K. was supported by Polish NSC via grant DEC-2013/10/E/ST9/00662.
V.\,M. was supported by Polish NSC grant 2016/22/E/ST9/00061.

\bsp	
\label{lastpage}

\bibliography{references}

\end{document}